\documentclass[letterpaper]{jpconf}
\usepackage{graphicx}
\begin{document}

\title{Numerical study of hot strongly interacting matter}

\author{P. Petreczky}

\address{Physics Department, Brookhaven National Laboratory, 
         Upton, NY 11973, USA}

\begin{abstract}
I review recent progress in study of strongly interacting matter
at high temperatures using Monte-Carlo simulations in lattice QCD.
\end{abstract}

\section{Introduction}
It is expected that strongly interacting matter undergoes a transition in some temperature
interval from hadron gas to deconfined state also called the quark gluon plasma (QGP) \cite{gros81}. 
Creating deconfined medium in a laboratory is 
the subject of the large experimental program at RHIC 
\cite{nagle} and  LHC \cite{salgado}.
Attempts to study QCD thermodynamics on the lattice go back to the early 80's when lattice
calculations in $SU(2)$ gauge theory provided the first rigorous theoretical evidence for
deconfinement \cite{kuti81,mclerran81,engels81}. The problem of calculating thermodynamic 
observables in pure
gluonic theory was solved in 1996 \cite{boyd}, while simulations involving dynamical quarks
were limited to large quark masses and had no control over the discretization errors 
\cite{milc_old_eos,wilson_old,p4_old}.
Lattice calculations of QCD thermodynamics with
light dynamical quarks remained challenging until recently. During the past six years calculations
with light $u,d$ quarks have been performed using improved staggered fermion actions 
\cite{milc_Tc,fodor05,our_Tc,fodor06,milc_eos,our_eos,eos005,hotQCD,fodor09,fodor10,fodor10eos,wwnd10,hotqcd2}.
There was also progress in lattice QCD calculations at non-zero temperature using Domain Wall
and Wilson fermion formulations \cite{prasad,fodor11,whotqcd}. However, due to much larger computational costs the
corresponding results are far less extensive.

To get reliable predictions from lattice QCD the lattice spacing $a$ should be sufficiently small
relative to the typical QCD scale, i.e. $\Lambda_{QCD} a \ll 1$. For staggered fermions, discretization
errors go like ${\cal O}( (a \Lambda_{QCD})^2)$ but discretization errors due to flavor symmetry
breaking turn out to be numerically quite large and dominate the cutoff dependence of thermodynamic
quantities at low temperatures.
To reduce these errors one has to use improved staggered fermion actions with so-called fat links
\cite{orginos}. At high temperature the dominant discretization errors go like $(a T)^2$ and therefore
could be very large. Thus it is mandatory to use improved discretization schemes, which improve the
quark dispersion relation and eliminate these discretization errors. Lattice fermion actions 
used in numerical calculations typically implement some version of fat links as well as improvement
of quark dispersion relation and are referred to as $p4$, $asqtad$, $HISQ/tree$ and $stout$. 

In this contribution I am going to discuss lattice QCD calculations on the equation of state, 
study of deconfinement aspects of the QCD transition, 
including color screening and fluctuations of conserved charges, determination of the chiral 
transition temperature, as well as calculations of temporal and spatial meson correlation functions.

\section{Equation of State}
\begin{figure}[ht]
\includegraphics[width=7.3cm]{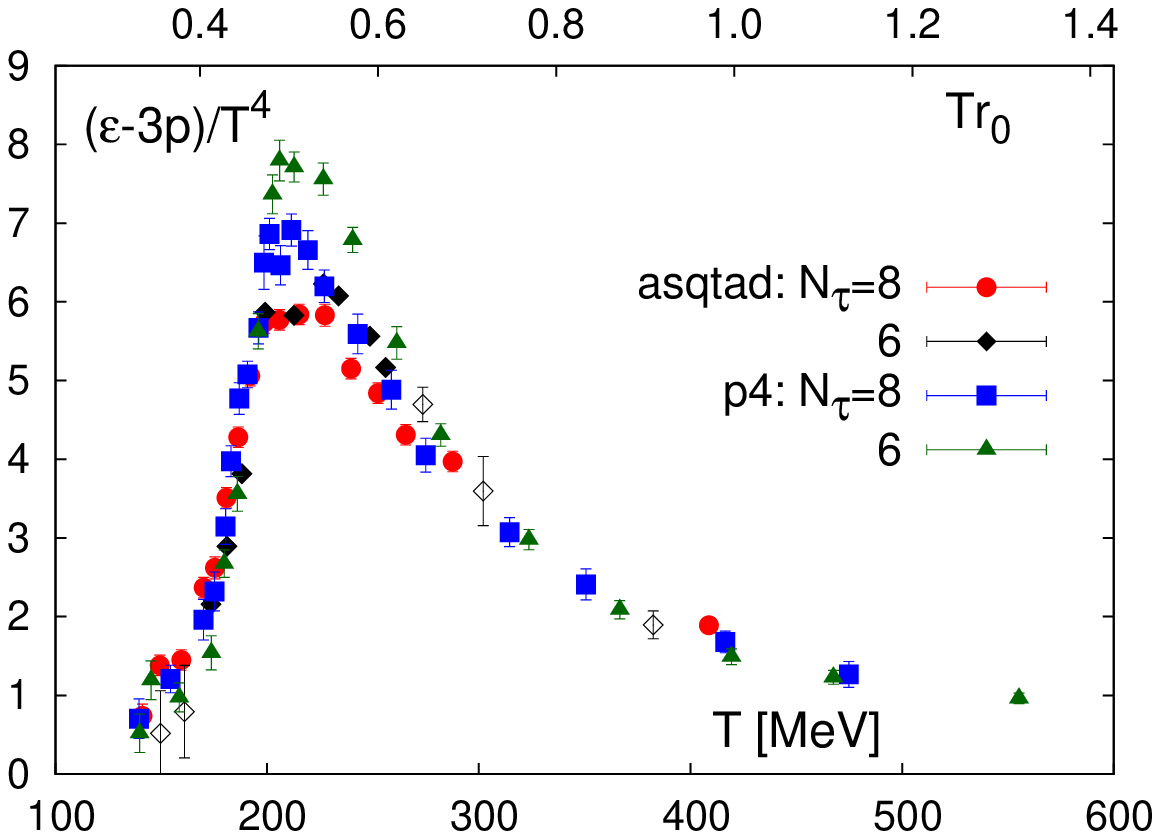} 
\includegraphics[width=7.3cm]{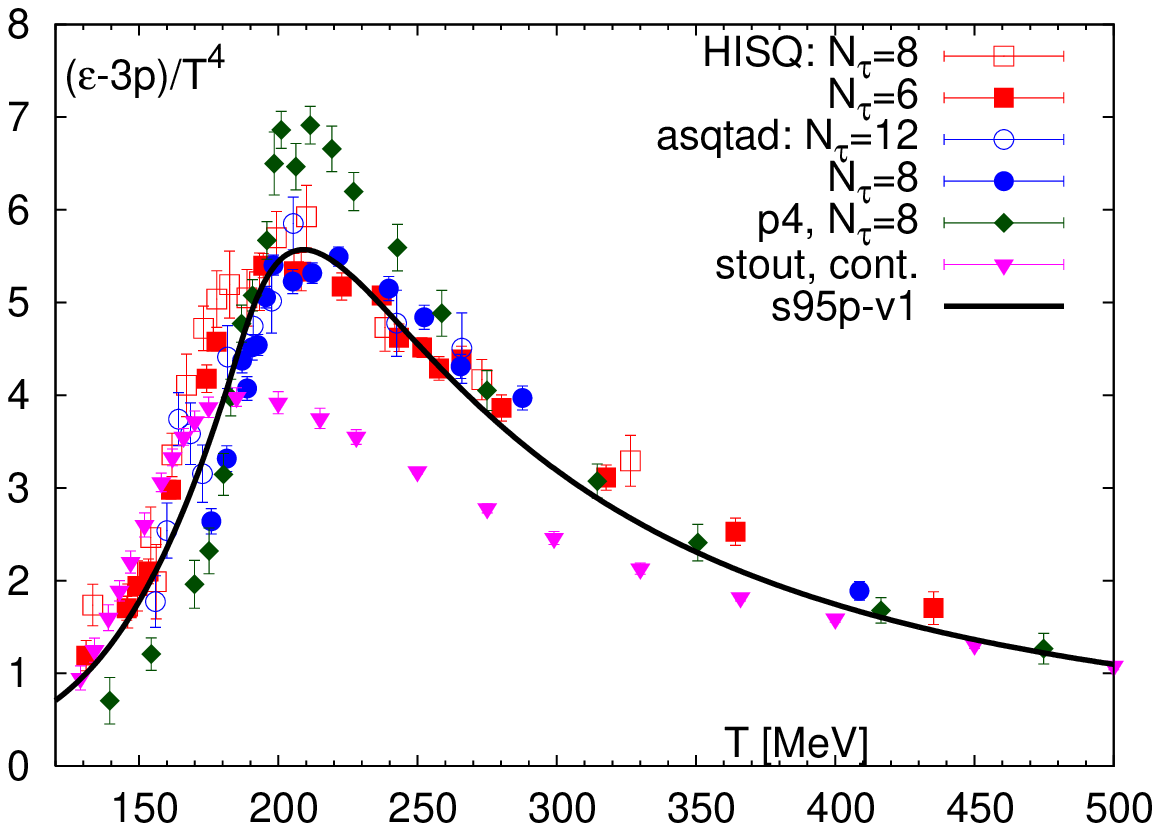}
\vspace*{-0.3cm}
\caption[]{
The interaction measure calculated with $m_l=m_s/10$ and $p4$ and $asqtad$ actions \cite{hotQCD} (left)
and with $m_l=m_s/20$ for $p4$ action \cite{eos005} as well as with $HISQ/tree$ and $asqtad$ actions \cite{latproc}.
Also shown in the figure are the continuum estimates obtained with $stout$ action and the parametrization
based on hadron resonance gas (HRG) model \cite{pasi}.
}
\label{fig:e-3p}
\end{figure}

The equation of state has been calculated using different improved staggered fermion actions
$p4$, $asqtad$, $stout$ and $HISQ/tree$. 
In the lattice calculations of the equation of state and many other quantities
the temperature is varied by varying the lattice spacing at fixed value of the temporal extent $N_{\tau}$.
The temperature $T$ is given by the lattice spacing and the temporal extent, $T=1/(N_{\tau} a)$. Therefore
taking the continuum limit corresponds to $N_{\tau} \rightarrow \infty$ at the fixed physical volume.
The calculation
of thermodynamic observables proceeds through the calculation of the trace of the energy momentum tensor,
$\epsilon -3 p$, also
known as trace anomaly or interaction measure. This is due to the fact that this quantity can be expressed in
terms of expectation values of local gluonic and fermionic operators, (see e.g. Ref. \cite{hotQCD}).  
Different thermodynamic observables can be obtained from the interaction measure through integration of
the trace anomaly
\footnote{A somewhat different approach was used in Ref. \cite{fodor10eos}}. 
The pressure can be written as
\begin{equation}
\displaystyle
\frac{p(T)}{T^4}-\frac{p(T_0)}{T_0^4}=\int_{T_0}^T \frac{dT'}{T'^5} (\epsilon-3 p).
\end{equation}
The lower integration limit $T_0$ is chosen such that the pressure is exponentially small there.
Furthermore, the entropy density can be written as $s=(\epsilon+p)/T$. Since the interaction measure 
is the basic thermodynamic observable in the lattice calculations it is worth to discuss its properties
more in detail. In Fig. \ref{fig:e-3p} (left panel) I show 
the results of calculation with $p4$ and $asqtad$ actions using $N_{\tau}=6$ and $8$ lattices and light
quark masses $m_l=m_s/10$, where $m_s$ is the physical strange quark mass. These calculations correspond
in the continuum limit to the pion mass of $220$MeV and $260$MeV for $p4$ and $asqtad$ respectively.
The interaction measure shows a rapid rise in the transition region and after reaching a peak at temperatures
of about $200$MeV decreases. Cutoff effects (i.e. $N_{\tau}$ dependence) 
appears to be the strongest around the peak region and decrease at high temperatures. 
For temperatures $T<270$MeV calculations with
$p4$ and $asqtad$  actions have been extended to smaller quark masses, $m_l=m_s/20$, that correspond to
the pion mass of about $160$MeV in the continuum limit \cite{eos005,latproc}.
It turns out that the quark mass dependence is negligible for $m_l < m_s/10$.  
Furthermore, for $astqad$  action calculations have been extended to $N_{\tau}=12$ lattices \cite{latproc}.
The trace anomaly was calculated with $HISQ/tree$ action on lattices with temporal extent $N_{\tau}=6$ and
$8$ and $m_l=m_s/20$ \cite{latproc} (corresponding to $m_{\pi}=160$MeV in the continuum limit). Finally, calculation
of the trace anomaly and the equation of state was performed with $stout$ action using $N_{\tau}=4,~6,~8,~10$ and
$12$ and physical light quark masses \cite{fodor10eos}. Using the lattice data from $N_{\tau}=6,~8,~10$ 
a continuum estimate for different quantities was given \cite{fodor10eos}. In  Fig. \ref{fig:e-3p} (right
panel) the results of different lattice calculations of $\epsilon-3p$ are summarized corresponding to the pion
masses close to the physical value. I also compare the lattice results with the parametrization s95p-v1 
of $\epsilon-3p$.
This parametrization combines lattice QCD results of Refs. \cite{our_eos,hotQCD} at high temperatures
with hadron resonance gas model (HRG) at low temperatures ($T<170$MeV) \cite{pasi}.
At low temperatures there is a fair agreement between the results obtained with $stout$ action and the results
obtained with $HISQ/tree$ action as well as with $asqtad$ action for $N_{\tau}=12$. All these lattice results
are slightly above the HRG curve. There is a big difference in the height of the peak in the calculations
obtained with $asqtad$ and $HISQ/tree$ actions and the continuum estimate for $stout$ action. The $p4$ action
has large cutoff effects at low temperatures and in the peak region. As the result the $p4$ results are higher
in the peak region and fall 
below the HRG curve at low temperatures. When cutoff effects in the hadron spectrum are taken into account
in the HRG model a good agreement between the $p4$ data and HRG result can be achieved \cite{pasi}.
Since the dominant cutoff effects of order $a^2 T^2$  are eliminated, the $N_{\tau}$-dependence is 
expected to be small for $p4$, $asqtad$ and $HISQ/tree$ action at high temperatures.
We see that for $T>250$MeV all these lattice actions lead to similar results. At temperatures above
$350$MeV we also see a good agreement with the $stout$ results. 

\begin{figure}[ht]
%\hspace*{-0.8cm}
\includegraphics[width=7.5cm]{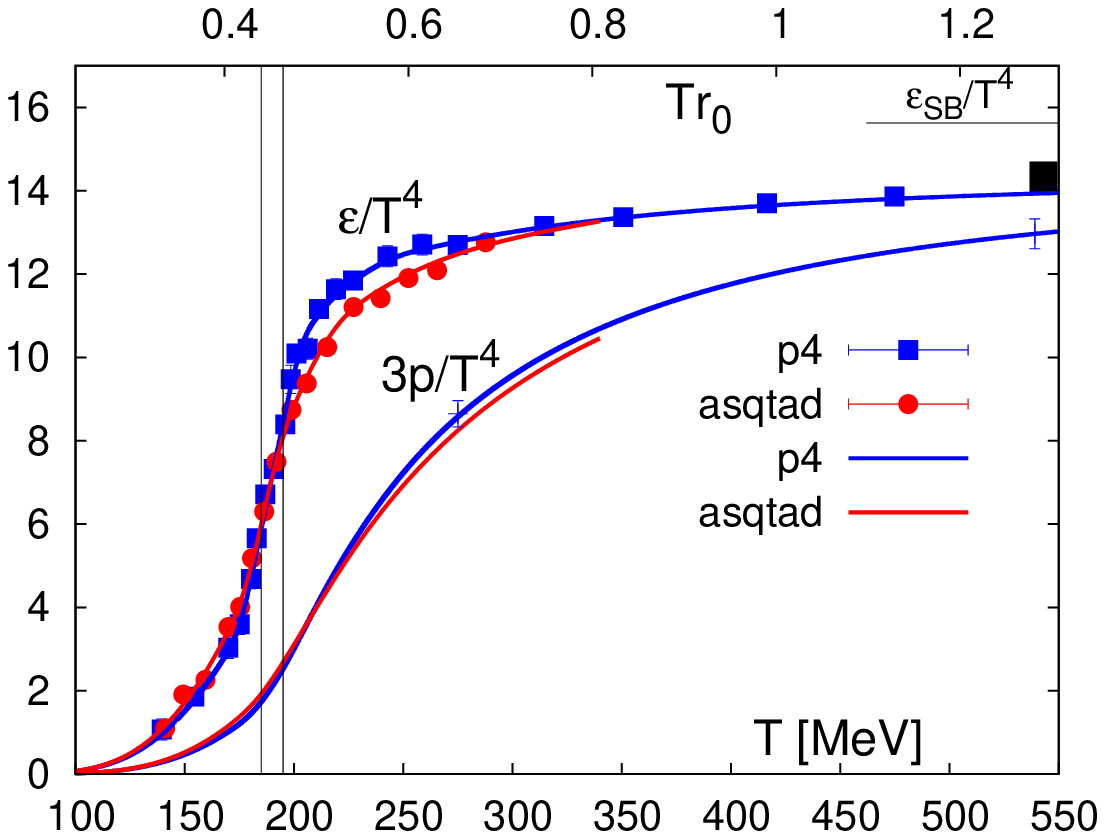}
\includegraphics[width=7.5cm]{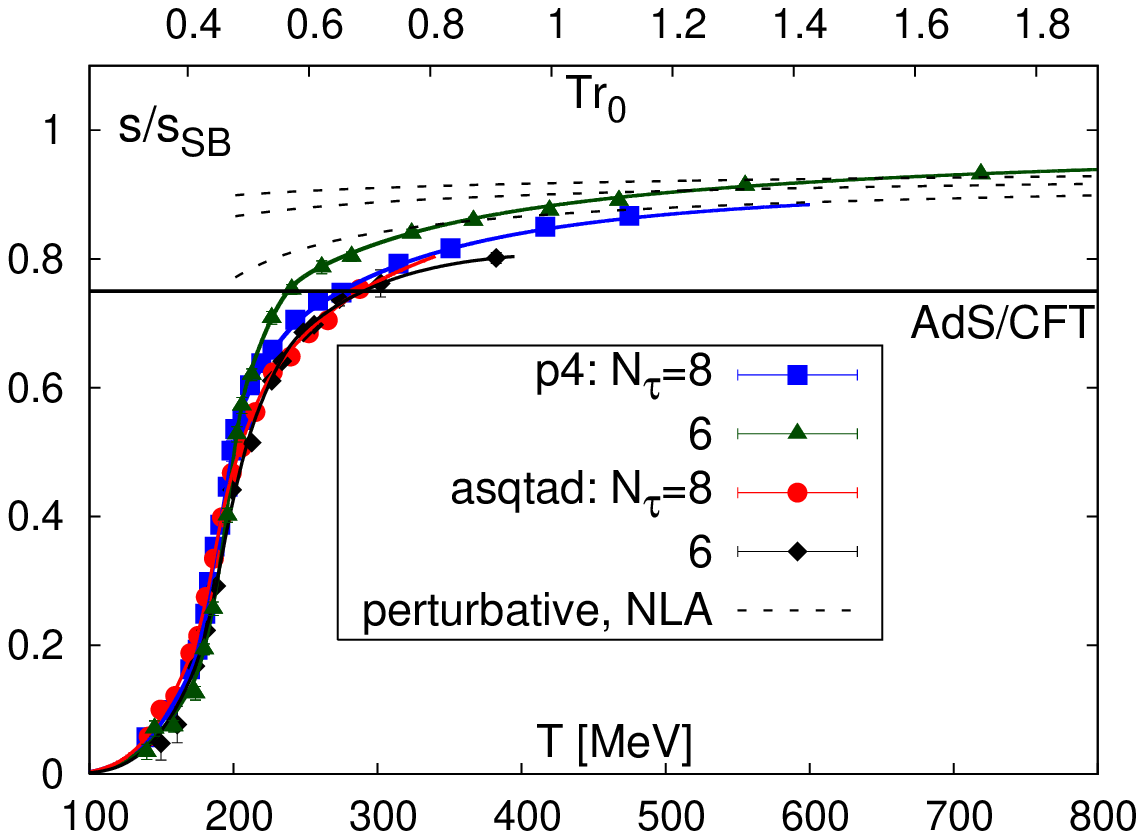}
\vspace*{-0.3cm}
\caption[]{
The energy density and the pressure as function of the temperature (left), and the entropy density divided
by the corresponding ideal gas value (right). The dashed lines in the right panel correspond to the resummed
perturbative calculations while the solid black line is the AdS/CFT result.
}
\label{fig:eos}
\end{figure}

The pressure, the  energy density and the entropy density are shown in Fig. \ref{fig:eos}. The energy density
shows a rapid rise in the temperature region $(170-200)$ MeV and quickly approaches about $90\%$ of the ideal gas
value. The pressure rises less rapidly but at the highest temperature it is also only about $15\%$ below the ideal
gas value. In the previous calculations with the $p4$ action it was found that the pressure and the energy density
are below the ideal gas value by about $25\%$ at high temperatures \cite{p4_old}.
Possible reason for this larger deviation could
be the fact that the quark masses used in this calculation were fixed in units of temperature instead 
being tuned to give constant meson masses as lattice spacing is decreased. As discussed
in Ref. \cite{fodor04} this could reduce the pressure by $10-15\%$ at high temperatures. 
In Fig. \ref{fig:eos} I also show the entropy density divided by the corresponding ideal gas value
and compare the results of lattice calculations with resummed perturbative calculation \cite{blaizot,blaizot1}
as well as with the predictions from AdS/CFT correspondence for the strongly coupled regime \cite{gubser98}.
The later is considerably below the lattice results. Note that the pressure, the energy density 
and the trace anomaly have also been recently discussed in the framework of 
resummed perturbative calculations which seem to agree with lattice data quite well at
high temperatures\cite{mike}.

The differences between the $stout$ action and the $p4$ and $asqtad$ actions for the trace anomaly
translates into the differences in the pressure and the  energy density. 
In particular, the energy density is about 20\% below the ideal gas limit for the $stout$ action. 

\section{Deconfinement : Fluctuations of conserved charges}
\begin{figure}[ht]
\includegraphics[width=7cm]{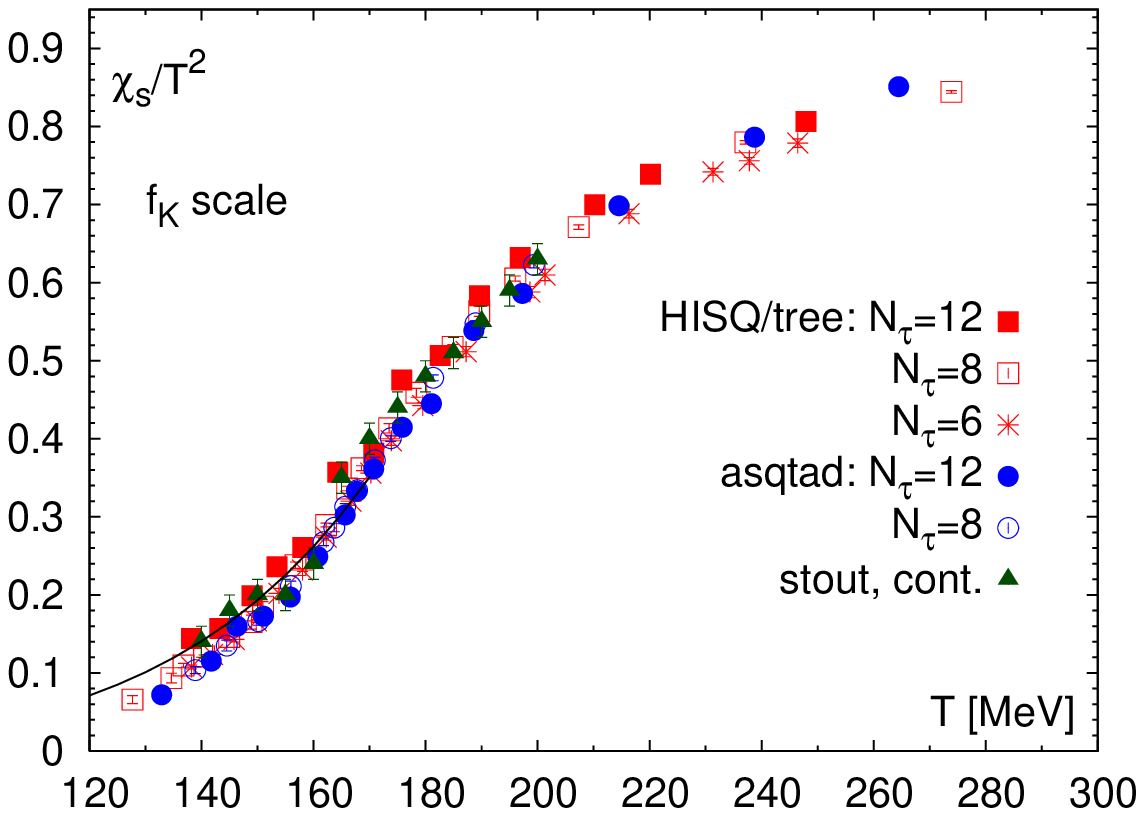}
\includegraphics[width=7cm]{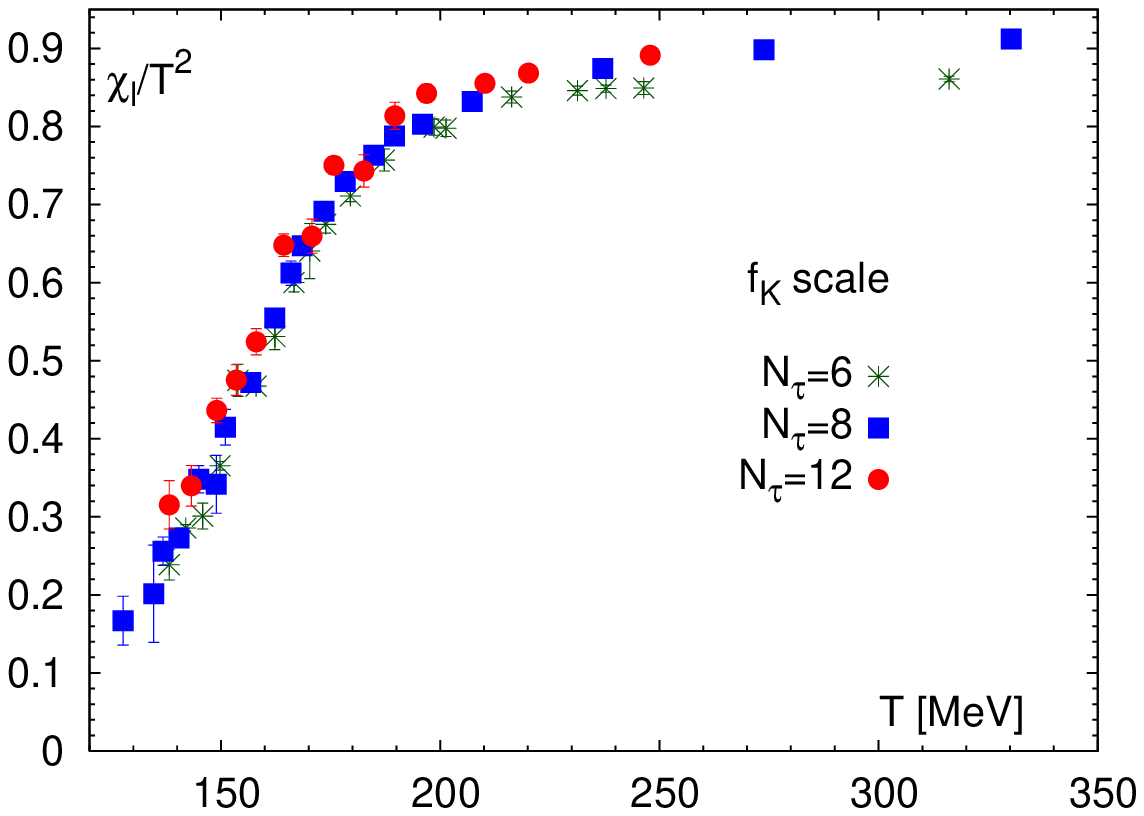}
\caption{Strangeness fluctuation calculated with $asqtad$ and $HISQ/tree$ actions
\cite{hotqcd2} and compared with the continuum estimate obtained with stout action \cite{fodor10} as well as
the prediction of the HRG model shown as a black line (left panel). In the right panel the light quark number
fluctuations are shown for $HISQ/tree$ action \cite{hotqcd2}.
}
\label{fig:chis}
\vspace*{-0.2cm}
\end{figure}

Due to the infamous sign problem lattice QCD Monte-Carlo simulations are not possible at non-zero
quark chemical  potentials.
The pressure at non-zero chemical potentials can be evaluated using Taylor expansion. The 
Taylor expansion can be set up in terms of quark chemical potentials or in terms of chemical
potentials corresponding to baryon number $B$, electric charge $Q$ and strangeness $S$ of hadrons.
The expansion coefficients in quark chemical potential $\chi_{uds}^{jkl}$ and 
the hadronic ones $\chi_{BQS}^{jkl}$ are related to each other \cite{our_fluct}.
While Taylor expansion can be used to study the physics at non-zero baryon density they are interesting on 
their own right as they are related to the fluctuations of conserved charges. The later are
sensitive probes of deconfinement. 
This is because fluctuation of conserved charges are sensitive to the underlying degrees of freedom
which could be hadronic or partonic. 
Fluctuations of conserved charges have been studied using different staggered 
actions \cite{milc_Tc,hotQCD,fodor10,our_fluct,fodor_fluct}. 
As an example in  Fig. \ref{fig:chis}
I show the quadratic strangeness fluctuation as the function of the temperature. 
Fluctuations are suppressed at low temperatures
because conserved charges are carried by massive strange hadrons (mostly kaons). 
They are well described
by Hadron Resonance Gas (HRG) model at low temperatures. 
Strangeness fluctuations rapidly grow in the transition region of
$T=(160-200)$ MeV as consequence of deconfinement. At temperatures $T>250$ MeV strangeness fluctuations are
close to unity that corresponds to the  massless ideal quark gas. Similar picture can seen in other fluctuations, 
in particular for light quark number fluctuation, $\chi_l$, which is also show in Fig. \ref{fig:chis}. 
The more rapid rise of $\chi_l$ in the transition region is presumably a quark mass effect.

\begin{figure}[ht]
\includegraphics[width=7cm]{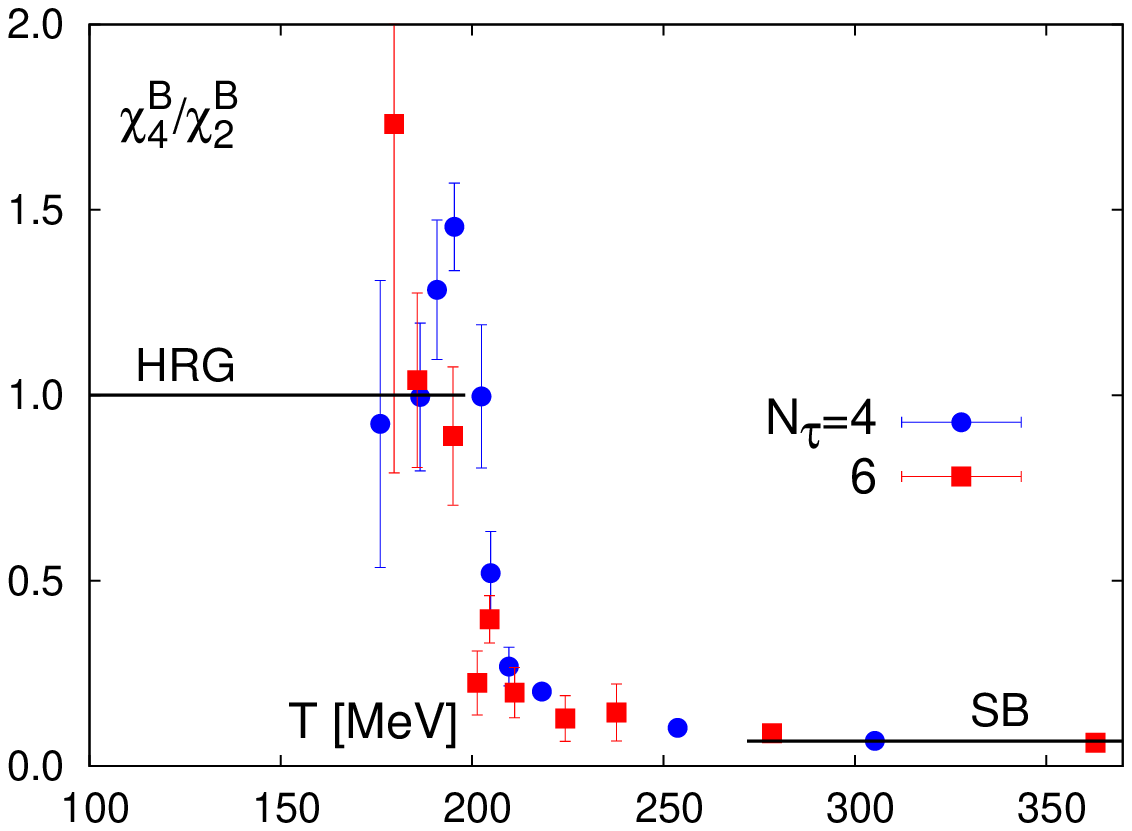}
\includegraphics[width=7cm]{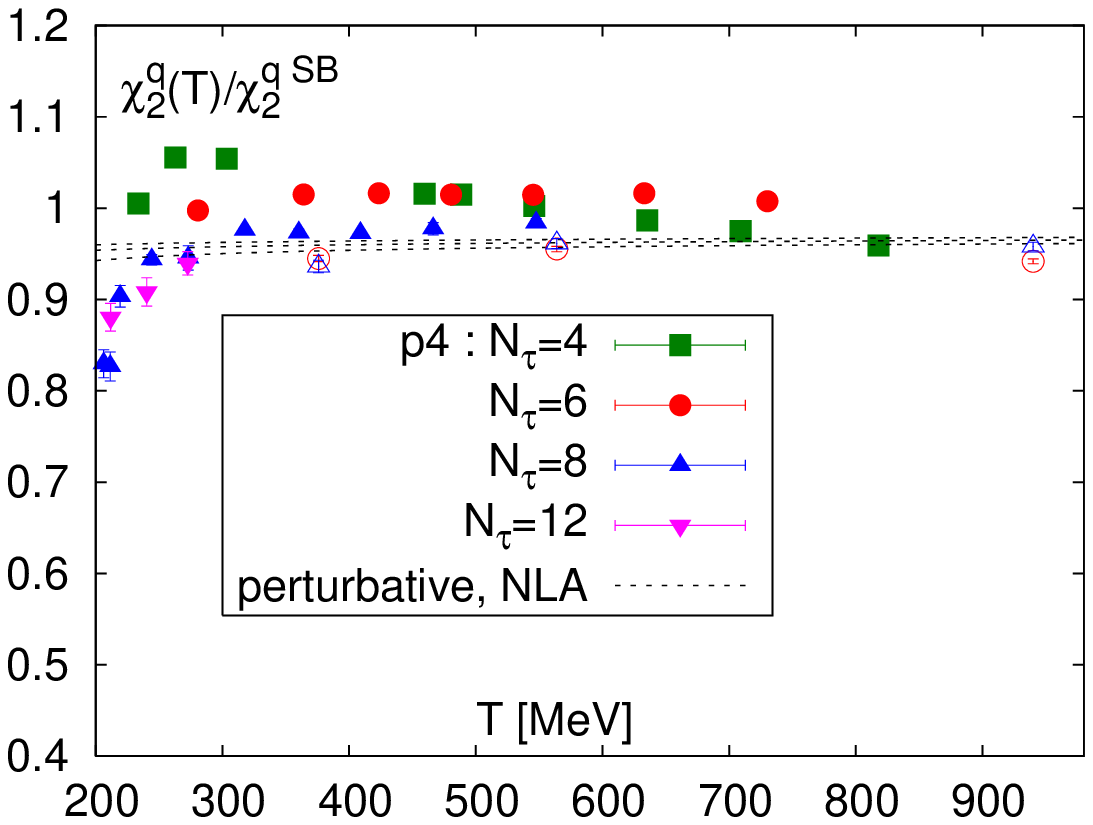}
\vspace*{-0.3cm}
\caption[]{
The Kurtosis of baryon number (left) and the quark susceptibility
compared with resummed perturbative calculations (right). 
The open symbols on the right plot correspond to $asqtad$ results 
for strange quark number susceptibility \cite{milc_Tc}.
The resummed perturbative
results were obtained in next-to-leading approximation (NLA) \cite{rebhan_sewm02}.
}
\label{fig:chi_highT}
\vspace*{-0.5cm}
\end{figure}

The transition from hadronic to quark degrees of freedom can be particularly well seen in the temperature
dependence of the Kurtosis, which is the ratio of quartic to quadratic fluctuations. This quantity can
be also measured experimentally. In Fig. \ref{fig:chi_highT} I show the Kurtosis
of the baryon number as a function of the temperature. At low temperatures it is temperature independent
and is close to unity in agreement with the prediction of the HRG model. In the transition
region it sharply drops from the hadron resonance gas value to the value corresponding to an ideal gas
of quarks. Since the fluctuations of conserved charges are so well described by ideal quark gas it is interesting
to compare the lattice results for quark number fluctuations with resummed perturbation theory. 
Quadratic quark number fluctuations are also called quark number susceptibilities and have been calculated
in resummed perturbation theory \cite{blaizot01,rebhan_sewm02,munshi}. The comparison of the resummed perturbative
results with lattice calculations performed with $p4$ action \cite{progress} 
for the quark number susceptibility divided
by the ideal gas value is shown in Fig. \ref{fig:chi_highT}. The figure shows a very good agreement
between the lattice calculations and resummed perturbative results. 
%It is also interesting to compare findings
%of the lattice calculations with the results obtained in strongly coupled gauge theories using AdS/CFT 
%correspondence. The result from AdS/CFT calculations shown in the figure as the solid black line is significantly
%below the lattice results \footnote{The conserved charges considered in these calculations are not exactly the
%quark numbers, see discussion in Ref. \cite{my_qm09}.}.
Let me finally note that the discretization errors
in the lattice calculations of the fluctuations are small at high temperatures when calculated with
the $p4$ action. The same is true for $asqtad$ action \cite{milc_Tc,hotQCD}. 

Fluctuations of conserved charges have also been  calculated using gauge/string duality in Refs. \cite{vs1,vs2}.
These calculations could reproduce several features of the lattice data.

\section{Deconfinement : color screening}

\begin{figure}
\includegraphics[width=8cm]{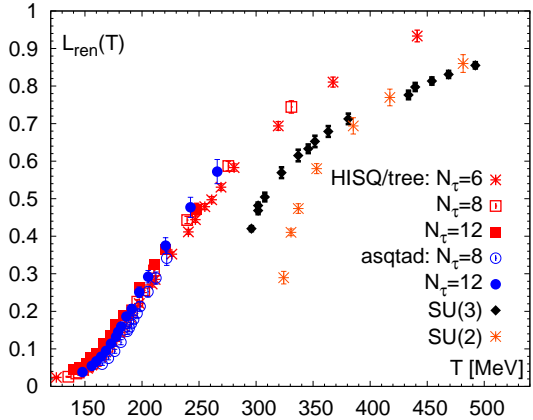}
\caption{The Polyakov loop as function of the temperature in pure gauge theory and 2+1 flavor QCD.}
\label{fig:lren}
\end{figure}

One of the most prominent feature of the quark gluon plasma is the presence of chromoelectric (Debye) screening.
The easiest way to study chromoelectric screening is to calculate the Polyakov loop.
The Polyakov loop is an order parameter for the deconfinement transition in pure gauge theory,
which is governed by $Z(N)$ symmetry. For QCD this symmetry is explicitly broken
by dynamical quarks. There is no obvious reason for the Polyakov loop
to be sensitive to the singular behavior close to the chiral limit although speculations
along these lines have been made \cite{speculations}. The Polyakov loop 
is related to the screening properties of the medium and thus to deconfinement.
After proper renormalization, the square of the Polyakov loop characterizes the
long distance behavior of the static quark anti-quark free energy; it 
gives the excess in free energy needed to screen two well-separated color
charges. The renormalized Polyakov loop has been studied in the past in 
pure gauge theory \cite{okacz02,digal03} as well as in QCD with two 
\cite{okacz05}, three \cite{kostya1} and  two plus one flavors 
\cite{our_eos,hotQCD}. The renormalized Polyakov loop, calculated on lattices 
with temporal extent $N_\tau$, is obtained from the bare Polyakov 
\begin{eqnarray}
&
\displaystyle
L_{ren}(T)=z(\beta)^{N_{\tau}} L_{bare}(\beta)=
z(\beta)^{N_{\tau}} \left<\frac{1}{3}  {\rm Tr } W(\vec{x}) \right >,~~W(\vec{x})=
\prod_{x_0=0}^{N_{\tau}-1} U_0(x_0,\vec{x}),
\end{eqnarray}
where $U_0=\exp(iga A_0)$ denotes the temporal gauge link and $z(\beta)$ 
is the renormalization constant determined from the $T=0$ static potential \cite{hotqcd2}.
The numerical results for the renormalized Polyakov loop for the $HISQ/tree$ action are
shown in  Fig.~\ref{fig:lren}. As one can see from the figure the cutoff 
($N_{\tau}$) dependence of the renormalized Polyakov loop is small. We also compare
our results with the continuum extrapolated $stout$ results \cite{fodor10} and the 
corresponding results in pure gauge theory \cite{okacz02,digal03}. 
We find good agreement between our results and the $stout$ results.
We also see that in the vicinity of the transition temperature the behavior of the renormalized
Polyakov loop in QCD and in the pure gauge theory is quite different.

Further insight on chromoelectric screening can be gained by studying
the singlet free energy of static quark anti-quark pair (for 
reviews on this see Ref. \cite{mehard04,qgp09}),  which is expressed in terms of 
correlation function of temporal Wilson lines in Coulomb gauge
\begin{equation}
\exp(-F_1(r,T)/T)=\frac{1}{3} {\rm Tr} \langle W(r) W^{\dagger}(0) \rangle.
\end{equation}
Instead of using the Coulomb gauge the singlet free energy can be defined in gauge invariant manner by
inserting a spatial gauge connection between the two Wilson lines. Using such definition the singlet free energy 
has been calculated in $SU(2)$ gauge theory \cite{baza08}.  
It has been found that the singlet free energy calculated this way is close to the result obtained in
Coulomb gauge \cite{baza08}. 
The singlet free energy turned out to be  useful to 
study quarkonia binding at high temperatures in potential models 
(see e.g. Refs. \cite{mocsy1,mocsy2,mocsy3,mocsy4,mocsy5}).  The singlet free energy 
also appears naturally in the perturbative
calculations of the Polyakov loop correlators at short distances \cite{plc_new}.

\begin{figure}
\includegraphics[width=7.5cm]{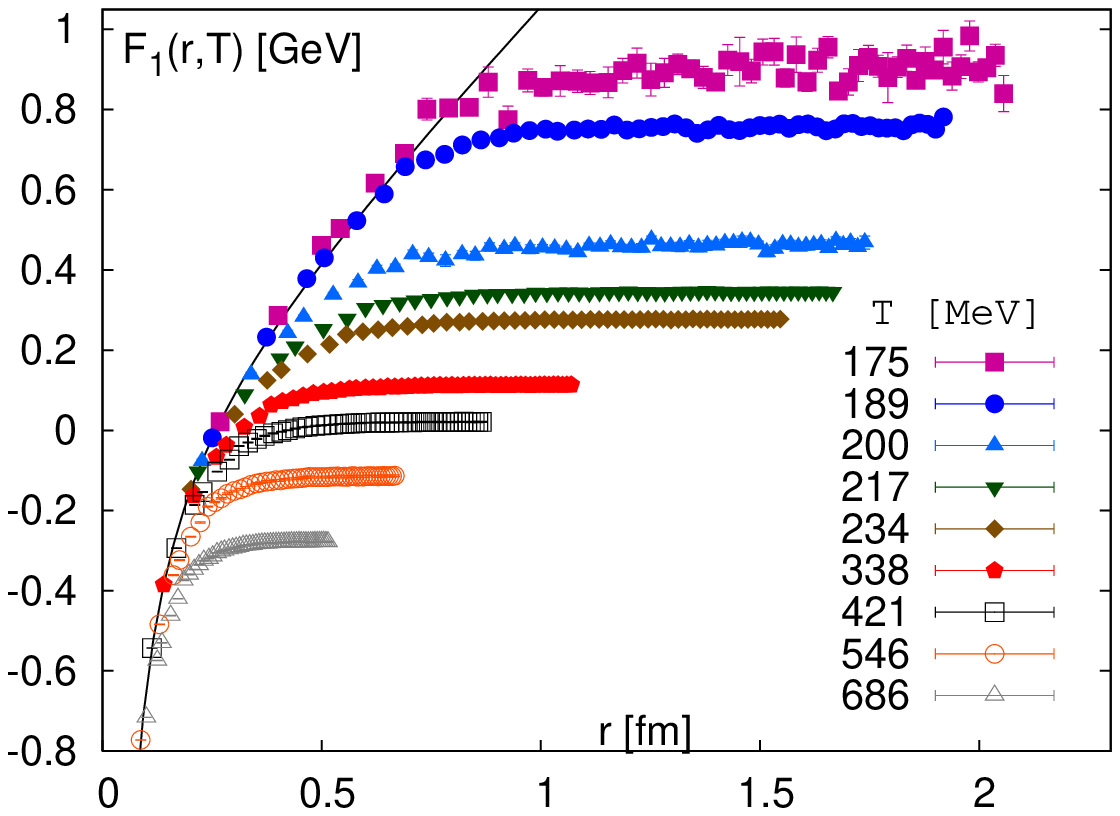}
\includegraphics[width=7.5cm]{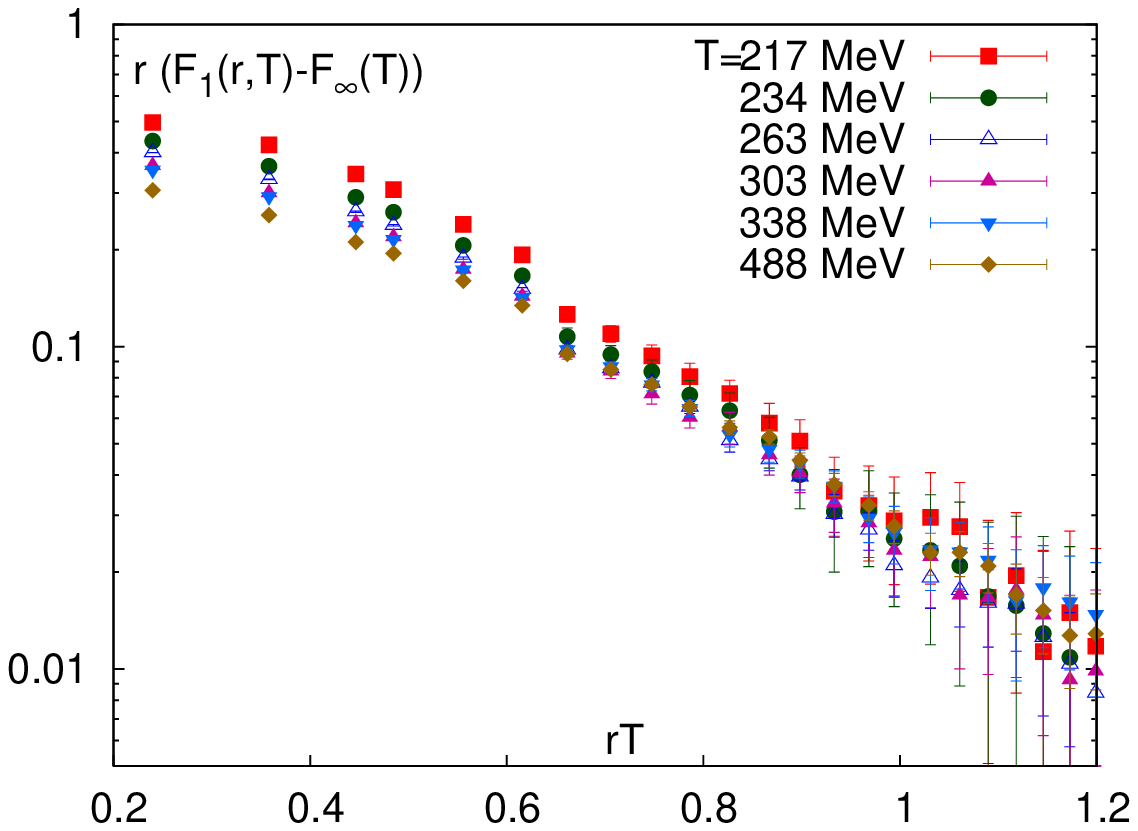}
\caption{The singlet free energy $F_1(r,T)$ calculated in Coulomb gauge on $16^3 \times 4$ lattices (left)
and the combination $F_1(r,T)-F_{\infty}(T)$ as function of $r T$ (right). The solid black line
is the parametrization of the zero temperature potential.}
\label{fig:f1} 
\end{figure}

The  singlet free energy was recently calculated in QCD with one strange quark and two light
quarks with masses corresponding to pion mass of $220$MeV on $16^3 \times 4$ lattices \cite{rbc_f1}. 
The numerical results are shown in Fig. \ref{fig:f1}.
At short distances the singlet free energy is temperature independent and coincides with the 
zero temperature
potential.  In purely gluonic theory the free energy grows linearly with the separation between 
the heavy quark and 
anti-quark in the confined phase. In presence of dynamical quarks the free energy is saturated 
at some finite value 
at distances of about $1$ fm due to string breaking \cite{mehard04,kostya1,okacz05}. 
This is also seen in Fig. \ref{fig:f1}. 
Above the deconfinement temperature the singlet free
energy is exponentially screened at sufficiently large 
distances \cite{okacz02,digal03} with the screening mass proportional to
the temperature , i.e.
\begin{equation}
F_1(r,T)=F_{\infty}(T)-\frac{4}{3}\frac{g^2(T)}{4 \pi r} \exp(-m_D(T) r), ~m_D \sim T.
\end{equation}
Therefore in Fig. \ref{fig:f1} I also show the combination $F_1(r,T)-F_{\infty}(T)$ 
as a function of $r T$. As one 
can see from the figure this function shows an exponential fall-off 
at distances $r T>0.8$. The fact that the slope is the same for all temperatures means
that $m_D \sim T$, as expected.

Let me finally note that contrary to the electro magnetic plasma the static chormomagnetic fields
are screened in QGP. This is due to the fact that unlike photons gluons interact with each other
(the stress tensor is non-linear in QCD). Magnetic screening is non-perturbative, i.e. it does not
appear at any finite order of pertubation theory. In lattice calculations chromomagnetic screening 
is studied either in terms of spatial Wilson loops \cite{sigma_s} or in terms of spatial gluon
propagators \cite{prop98,prop00,prop01}. The numerical results obtained so far show that the length
scale related to magnetic screening is larger than the one related to electric screening.

\section{Chrial transition}
The Lagrangian of QCD  has an approximate $SU_A(3)$ chiral symmetry. This symmetry is broken in the 
vacuum. The chiral symmetry breaking is signaled by  non-zero expectation value of the quark or chiral condensate,
$\langle \bar \psi \psi \rangle \neq 0$ in the massless limit.
This symmetry is expected to be restored at high temperatures and the quark condensate vanishes.
There is an explicit breaking of the chiral symmetry by the small value of $u,d$ and $s$ quark masses. 
While due to the relatively large strange quark mass ($m_s \simeq 100$ MeV) $SU_A(3)$ may not
be a very good symmetry its subgroup $SU_A(2)$ remains a very good symmetry and is relevant
for the discussion of the finite temperature transition in QCD. If the relevant symmetry is 
$SU_A(2)$ the chiral transition is expected to be second order for massless light ($u$ and $d$) quarks
belonging to the $O(4)$ universality class. Recent calculations with $p4$ action support this pictures \cite{MEoS}.
This also means that for non-zero light quark masses the transition must be a crossover. The later
fact seems to be supported by calculations in Ref. \cite{nature}.
The $U_A(1)$ symmetry is explicitly
broken in the vacuum by the anomaly but it is expected to be effectively restored at high temperatures
as non-perturbative vacuum fluctuations responsible for its breaking are suppressed at high temperatures.
If the  $U_A(1)$ symmetry is restored at the same temperature as the  $SU_A(2)$ symmetry the transition
could be first order \cite{pisarski81}. Recent calculations with staggered as well as with domain wall fermions
suggest that $U_A(1)$  symmetry gets restored at temperature that is significantly higher than the chiral
transition temperature \cite{prasad,mscr}.

For massless quark the chiral condensate vanishes at the critical temperature $T_c^0$ and is the order
parameter. Therefore in the lattice studies one calculates the chiral condensate and its derivative
with respect to the quark mass called the chiral susceptibility. For the staggered fermion formulation
most commonly used in the lattice calculations these quantities can be written as follows:
\begin{eqnarray}
\langle \bar \psi \psi \rangle_{q,x}&=&\frac{1}{4} \frac{1}{N_{\sigma}^3 N_{\tau}} 
{\rm Tr} \langle D_q^{-1} \rangle ,\\
\chi_{m,q}(T)&=& 
n_f \frac{\partial \langle \bar\psi \psi \rangle_{q,\tau}}{\partial m_l}
=\chi_{q, disc} + \chi_{q, con} ~~q=l,s, \label{susc}
\end{eqnarray}
where the subscript $x=\tau$ and $x=0$ will denote the expectation value
at finite and zero temperature, respectively.  
Furthermore, $D_q=m_q \cdot 1 + D$ is the fermion matrix in the canonical normalization and
$n_f=2$ and $1$ for light and strange quark.
In Eq. (\ref{susc} we made explicit that chiral susceptibility is the sum of connected 
and disconnected Feynman diagrams. The disconnected and connected contributions can be written
as
\begin{eqnarray}
\chi_{q, disc} &=&
{{n_f^2} \over 16 N_{\sigma}^3 N_{\tau}} \left\{
\langle\bigl( {\rm Tr} D_q^{-1}\bigr)^2  \rangle -
\langle {\rm Tr} D_q^{-1}\rangle^2 \right\}
\label{chi_dis} \; , \\
\chi_{q, con} &=&  -
{{n_f} \over 4} {\rm Tr} \sum_x \langle \,D_q^{-1}(x,0) D_q^{-1}(0,x) \,\rangle \; ,~~~q=l,s.
\label{chi_con}
\end{eqnarray}
The disconnected part of the light quark susceptibility describes the
fluctuations in the light quark condensate and is directly analogous to
the fluctuations in the order parameter of an $O(N)$ spin model. The
second term ($\chi_{q,con}$) arises from the explicit quark mass
dependence of the chiral condensate and is the expectation value of
the volume integral of the correlation function of the (isovector)
scalar operator $\bar{\psi}\psi$. Let me note that in the massless linit only
$\chi_{l,disc}$ diverges. 

\subsection{The temperature dependence of the chiral condensate}
\begin{figure}
\includegraphics[width=7cm]{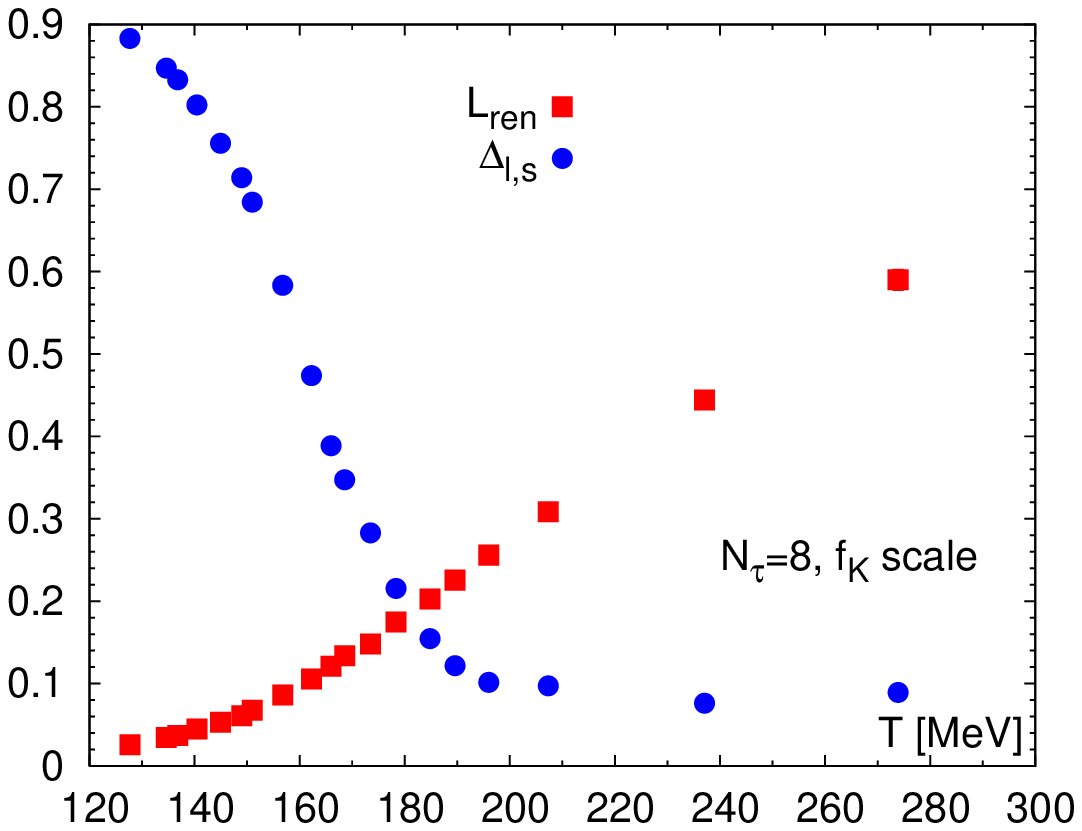}
\includegraphics[width=7cm]{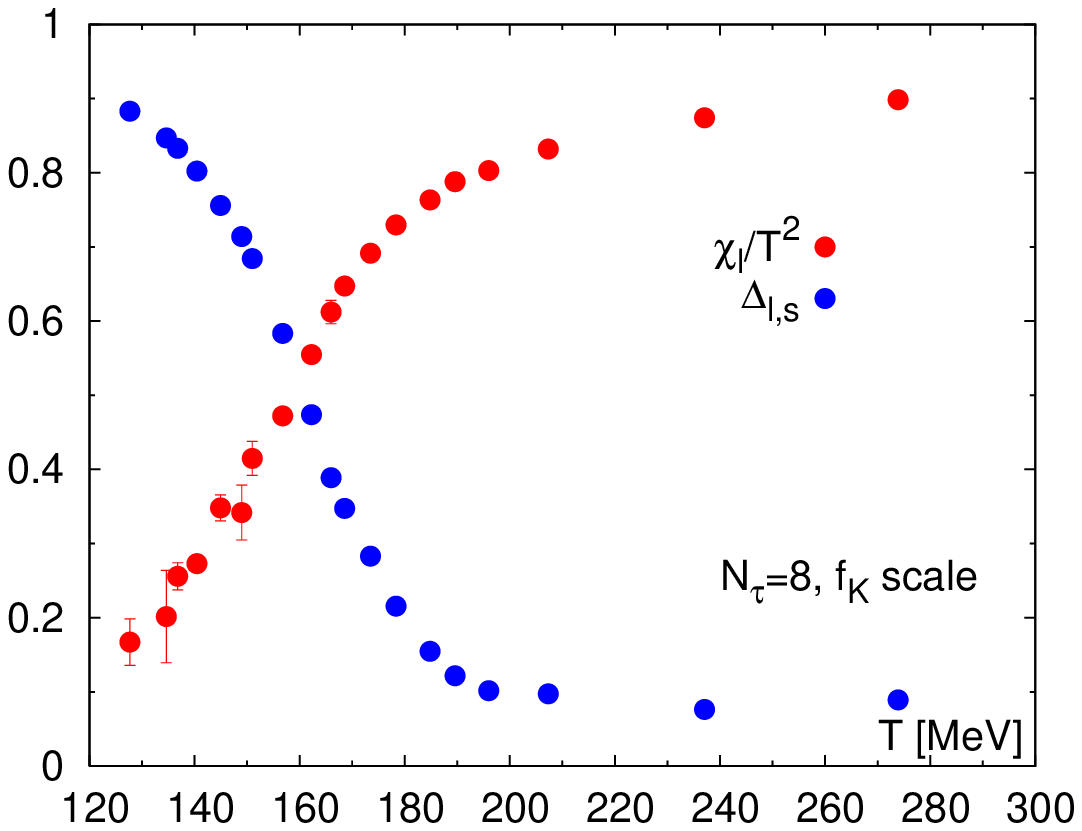}
\caption{The subtracted chiral condensate calculated with $HISQ/tree$ action on $N_{\tau}=8$ lattice
and compared to the renormalized Polyakov loop (left) and light quark number fluctuation (right).}
\label{fig:Delta}
\end{figure}

The chiral condensate needs a multiplicative, and also an additive renormalization if
the quark mass is non-zero. Therefore the subtracted chiral condensate is 
considered
\begin{equation}
\Delta_{l,s}(T)=\frac{\langle \bar\psi \psi \rangle_{l,\tau}-\frac{m_l}{m_s} \langle \bar \psi \psi \rangle_{s,\tau}}
{\langle \bar \psi \psi \rangle_{l,0}-\frac{m_l}{m_s} \langle \bar \psi \psi \rangle_{s,0}}.
\end{equation}
In Fig. \ref{fig:Delta} I show the results for $\Delta_{l,s}$ calculated with $HISQ/tree$ action and compared to
the renormalized Polyakov loop and light quark number fluctuation discussed in relation to the deconfining
transition. Interestingly the rapid decrease in the subtracted chiral condensate happens at temperatures
that are smaller than the temperatures where the Polyakov loop rises rapidly. On the other hand the rapid
change in $\chi_l$ and $\Delta_{l,s}$ happen roughly in the same temperature region.

\begin{figure}
\includegraphics[width=0.450\textwidth]{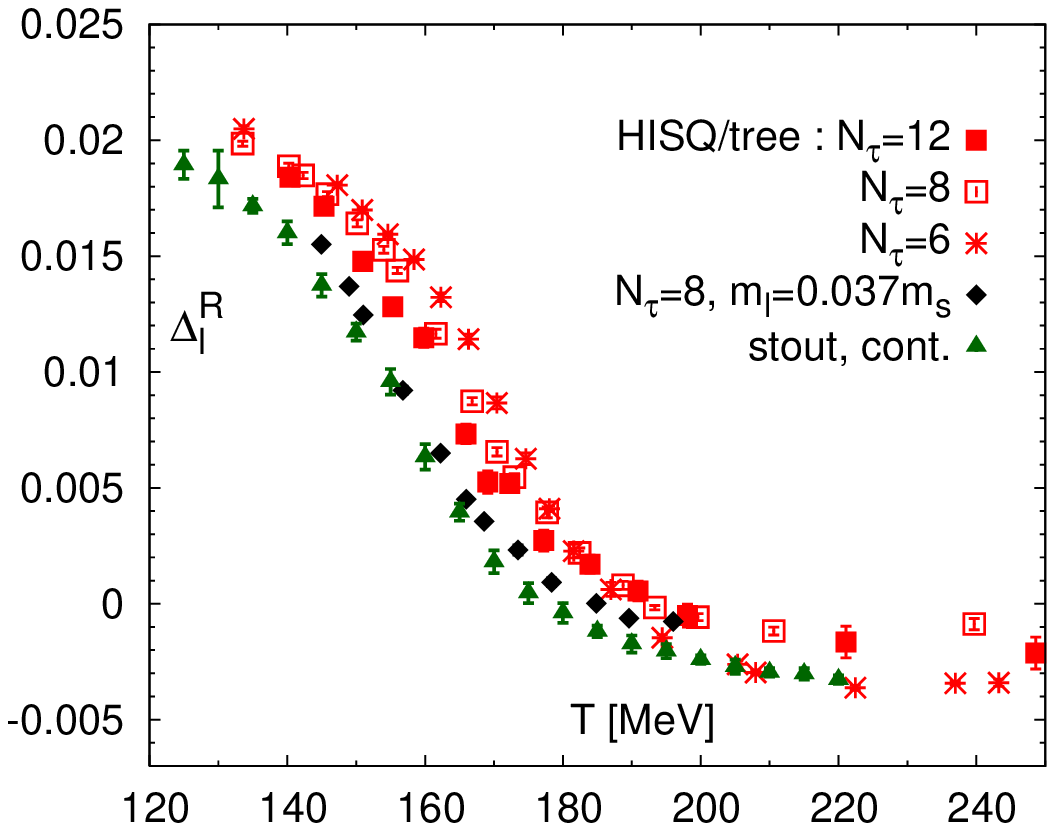}
\includegraphics[width=0.450\textwidth]{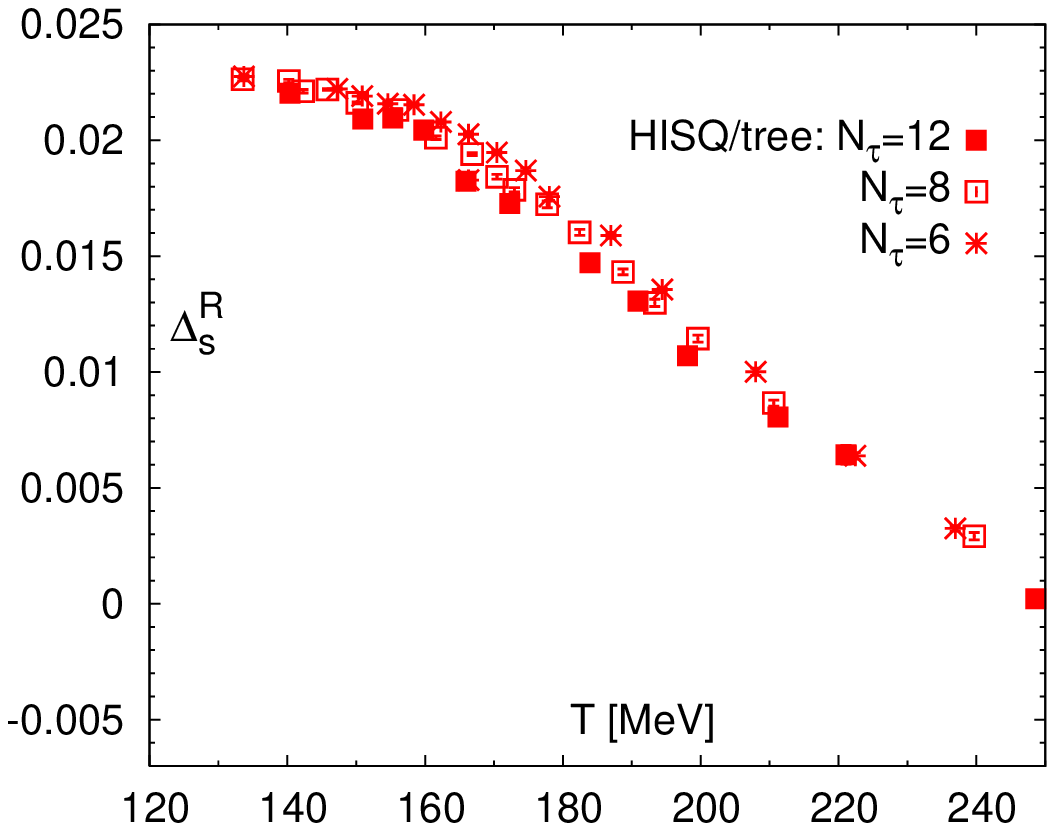}
\caption{The renormalized chiral condensate $\Delta_l^R$ for the
  $HISQ/tree$ action with $m_l/m_s = 0.05$ is compared to the $stout$ data.
  In the right panel, we show the renormalized strange quark
  condensate $\Delta_s^R$ for the $HISQ/tree$ action.
}
\label{fig:pbpR}
\end{figure}

Another way to get rid of the multiplicative and additive renormalization is to subtract the zero
temperature condensate and multiply the difference by the strange quark mass, i.e. consider the following
quantity
\begin{equation}
\Delta_q^R=d+n_f m_s r_1^4 ( \langle \bar \psi \psi \rangle_{q,\tau}-
\langle \bar \psi \psi \rangle_{q,0} ) ~ q=l,s .
\end{equation}
As before, $n_f=2$ for light quarks and $n_f=1$ for strange quarks, while $d$ is a normalization constant.
The factor $r_1^4$ was introduce to make the combination dimensionless. Here $r_1$ is the
scale parameter defined from the zero temperature static potential \cite{hotqcd2}.
It is convenient to choose the normalization constant to be the light quark condensate for $m_l=0$
multiplied by $m_s r_1^4$.  In Fig. \ref{fig:pbpR} the renormalized quark condensate is shown as function of
temperature for $HISQ/tree$ and $stout$ actions. We see a crossover behavior for temperature of $(150-160)$MeV,
where $\Delta_q^R$ drops by $50\%$. The difference between the $stout$ and $HISQ/tree$ results is a quark mass
effect. Calculations for $HISQ/tree$ action were performed for $m_{\pi}=160$ MeV, while the $stout$ calculations
were done for the physical quark mass. For a direct comparison with $stout$ results, we extrapolate the 
$HISQ/tree$ data in the light quark mass and also take care of the residual cutoff dependence in the $HISQ/tree$
data. This was done in Ref. \cite{hotqcd2} and the results are shown in the figure as black diamonds demonstrating
a good agreement between $HISQ/tree$ and $stout$ results.
Contrary to $\Delta_l^R$ the renormalized strange quark
condensate $\Delta_s^R$ shows only a gradual decrease over a wide temperature interval dropping by $50\%$
only at significantly higher temperatures of about $190$ MeV.

\subsection{The crial susceptibility}
For a true chiral phase transition the chiral susceptibility diverges at the transition temperature.
For physical value of the quark masses we expect to see a peak in the chiral susceptibility at certain
temperature that defines the crossover temperature. The chiral susceptibility also needs a multiplicative
and additive renormalization Therefore the following quantity is considered
\begin{equation}
\frac{\chi_R(T)}{T^4}=\frac{m_s^2}{T^4} \left( \chi_{m,l}(T)-\chi_{m,l}(T=0) \right).
\label{chiR}
\end{equation}
The numerical results for this quantity were presented in Ref. \cite{fodor06} for $stout$ action and
in Ref. \cite{hotqcd2} for the $HISQ/tree$ action. It was shown that the results of the two calculations
agree quite well up to small quark mass effects \cite{hotqcd2} \footnote{In Ref. \cite{fodor06} the light quark mass
was used instead of $m_s$ in Eq. (\ref{chiR}). In the comparison with was taken into account \cite{hotqcd2}.}.    

\subsection{O(N) scaling and the transition temperature}

In the vicinity of the chiral phase transition, the free energy
density may be expressed as a sum of a singular and a regular
part,
\begin{equation}
f = -\frac{T}{V} \ln Z\equiv f_{sing}(t,h)+ f_{reg}(T,m_l,m_s) \; .
\label{free_energy}
\end{equation}
Here $t$ and $h$ are dimensionless couplings that control deviations from
criticality. They are related to the temperature $T$ and the light quark mass $m_l$ as  
\begin{equation}
t = \frac{1}{t_0}\frac{T-T_c^0}{T_c^0} \quad , \quad 
h= \frac{1}{h_0} H \quad , \quad 
H= \frac{m_l}{m_s} \; ,
\label{reduced}
\end{equation}
where $T_c^0$ denotes the chiral phase transition temperature, 
{\it i.e.}, the transition temperature at $H=0$.  The scaling variables
$t$, $h$ are normalized by two parameters $t_0$ and $h_0$, which are
unique to QCD and similar to the low energy constants in the chiral
Lagrangian.  These need to be determined together with $T_c^0$. In the
continuum limit, all three parameters are uniquely defined, but depend
on the value of the strange quark mass.

The singular contribution to the free energy density is a homogeneous
function of the two variables $t$ and $h$. Its invariance under scale
transformations can be used to express it in terms of a single
scaling variable
\begin{equation}
z=t/h^{1/\beta\delta} = \frac{1}{t_0}\frac{T-T_c^0}{T_c^0} \left( \frac{h_0}{H} \right)^{1/\beta\delta}
 = \frac{1}{z_0}\frac{T-T_c^0}{T_c^0} \left( \frac{1}{H} \right)^{1/\beta\delta}
\label{eq:defz}
\end{equation}
where $\beta$ and $\delta$ are the critical exponents of the $O(N)$
universality class and $z_0 = t_0/h_0^{1/\beta\delta}$.
Thus, the dimensionless free energy density
$\tilde{f}\equiv f/T^4$ can be written as
\begin{equation}
\tilde{f}(T,m_l,m_s) = h^{1+1/\delta} f_f(z) + f_{reg}(T,H,m_s) \; ,
\label{scaling}
\end{equation}
where $f_f$ is the universal scaling function and the regular term $f_{reg}$ 
gives rise to scaling violations. This regular term 
can be expanded in a Taylor series around $(t,h)=(0,0)$.

It should be noted that the reduced temperature $t$ may depend on other
couplings in the QCD Lagrangian which do not explicitly break 
chiral symmetry. In particular, it depends on light and strange 
quark chemical potentials $\mu_q$, which in leading order enter only quadratically,
\begin{equation}
t = \frac{1}{t_0} \left( \frac{T-T_c^0}{T_c^0} + 
\sum_{q=l,s}\kappa_q\left(\frac{\mu_q}{T}\right)^2 +
\kappa_{ls} \frac{\mu_l}{T}\frac{\mu_s}{T} \right)
  \; .
\label{reduced2}
\end{equation}

The transition temperature can be defined as peaks in susceptibilities (response functions) that are 
second derivatives of the free energy density with respect to relevant parameters. Since there are
two relevant parameters we can define three susceptibilities:
\begin{equation}
\chi_{m,l}=\frac{\partial^2 \tilde f}{\partial m_l^2},~~
\chi_{t,l}=\frac{\partial^2 \tilde f}{\partial t \partial m_l },~~
\chi_{t,t}=\frac{\partial^2 \tilde f}{\partial t^2}.
\end{equation}
Thus three different  pseudo-critical temperatures $T_{m,l}$, $T_{t,l}$ and $T_{t,t}$ can be defined. 
In the vicinity of
the critical point the behavior of these susceptibilities is controlled by three universal
scaling function that can be derived from $f_f$. In the chiral limit $T_{m,l}=T_{t,l}=T_{t,t}=T_c^0$. 
There is, however, an additional complication for $O(N)$ universality class:  while $\chi_{m,l}$ and $\chi_{t,l}$
diverge at the critical point for $m_l \rightarrow 0$
\begin{equation}
\chi_{m,l} \sim m_l^{1/\delta - 1},~~~
\chi_{t,l} \sim m_l^{(\beta -1)/\beta\delta}, 
\label{peaks}
\end{equation}
$\chi_{t,t}$ is finite because $\alpha<0$ for $O(N)$ models ( $\chi_{t,t} \sim |t|^{-\alpha}$  ).
Therefore, one has to consider the third derivative of $\tilde f$ with respect to $t$ :
\begin{equation}
\chi_{t,t,t}=\frac{\partial^3 \tilde f}{\partial t^3}.
\end{equation}

In the vicinity of the critical point the derivatives with respect to $t$ can be estimated 
by taking the derivatives with respect to $\mu_l^2$, i.e.
the response functions $\chi_{t,l}$ and $\chi_{t,t,t}$ are identical to the first Taylor expansion coefficient
of the quark condensate and the sixth order expansion coefficient to the pressure, respectively. The former
controls the curvature of the transition temperature as function of the quark chemical potential $\mu_q$ and
was studied for $p4$ action using $N_{\tau}=4$ and $8$ lattices \cite{curvature}. The later corresponds to the sixth
order quark number fluctuation which is related to the deconfinement aspects of the transition.
The fact that this quantity is sensitive to the chiral dynamics points to a relation between  deconfining and chiral
aspects of the transition. 
In the following I discuss the determination of the transition temperature defined as
peak position of $\chi_{m,l}$, i.e. $T_c=T_{m,l}$. 

\subsection{Determination of the transition temperature}
The $O(N)$ scaling described in the above subsection can be used to determine the pseudo-critical temperature of
the chiral transition. For the study of the $O(N)$ scaling it is convenient to consider the dimensionless 
order parameter
\begin{equation}
M_b=m_s \frac{\langle \bar \psi \psi\rangle_l}{T^4}.
\end{equation}
The subscript "b" refers to the fact that this is a bare quantity since the additive UV divergence is not removed.
From the point of view of the scaling analysis this divergent term is just a regular contribution.
For sufficiently small quark mass and in the vicinity of the transition region we can write 
\begin{equation}
M_b(T,H) = h^{1/\delta} f_G(t/h^{1/\beta\delta}) + f_{M,reg}(T,H). 
\label{order_scaling}
\end{equation}
Here $f_G(z)$ is the scaling function related to $f_f$ and was calculated for $O(2)$ and $O(4)$ spin models
%\cite{Engels:2000xw,Toussaint:1996qr,Engels:1999wf,Engels:2001bq}.
\cite{EngelsO2,Toussaint,EngelsO4}.
The regular contribution can be parametrized as \cite{hotqcd2}
\begin{eqnarray}
f_{M,reg}(T,H) &=& a_t(T)  H   \nonumber \\
&=& \left( a_0 + a_1 \frac{T-T_c^0}{T_c^0} + a_2 \left(\frac{T-T_c^0}{T_c^0} \right)^2 \right) H.
\label{eq:freg}
\end{eqnarray}
Then we have the following behavior for the light chiral susceptibility
\begin{eqnarray}
\frac{\chi_{m,l}}{T^2} &=& \frac{T^2}{m_s^2}
\left( \frac{1}{h_0}
h^{1/\delta -1} f_\chi(z) + \frac{\partial f_{M,reg}(T,H)}{\partial H}
\right) \; , \nonumber \\
&&{\rm with}\;\; f_{\chi}(z)=\frac{1}{\delta} [f_G(z)-\frac{z}{\beta} f_G'(z)].
\label{eq:chiralsuscept}
\end{eqnarray}
One then performs a simultaneous fit to the lattice data for $M_b$ and $\chi_{m,l}$ treating
$T_c^0, t_0, h_0, a_0, a_1$ and $a_2$ as fit parameters \cite{hotqcd2}. This gives a good description
of the quark mass and temperature dependence of $\chi_{m,l}$ and allows to determine
accurately the peak position in $\chi_{m,l}$.
As an example in Fig. I show the $O(4)$ scaling fits for $N_{\tau}=8$ lattice data obtained with
$HISQ/tree$ action. 
The scaling fit works quite well. Similar results have been obtained for $N_{\tau}=6$ and $12$ as
well as for $asqtad$ action on $N_{\tau}=8$ and $12$ lattices \cite{hotqcd2}. Furthermore, scaling fits have been
performed assuming $O(2)$ universality class. The quality of these fits were similar to the $O(4)$ 
ones and the resulting
transition temperatures turned out to be the same within statistical errors \cite{hotqcd2}. 
Having determined $T_c$ for 
$HISQ/tree$  and $asqtad$ action for each $N_{\tau}$ a combined continuum extrapolation was performed using different
assumption about the $N_{\tau}$ dependence. of $T_c$ This analysis gave \cite{hotqcd2}:
\begin{equation}
T_c=(159 \pm 9) {\rm MeV}.
\end{equation}
The analysis also demonstrated that $HISQ/tree$ and $asqtad$ action give consistent results in the continuum limit.
The Budapest-Wuppertal collaboration found $T_c=147(2)(3)$MeV, $157(3)(3)$MeV and 
$155(3)(3)$MeV defined as peak position in $\chi_R$, inflection points in $\Delta_{l,s}$ 
and $\Delta_l^R$ respectively \cite{fodor10}. These agree with the above value within errors.
\begin{figure}[b]
\begin{center}
\includegraphics[width=7cm]{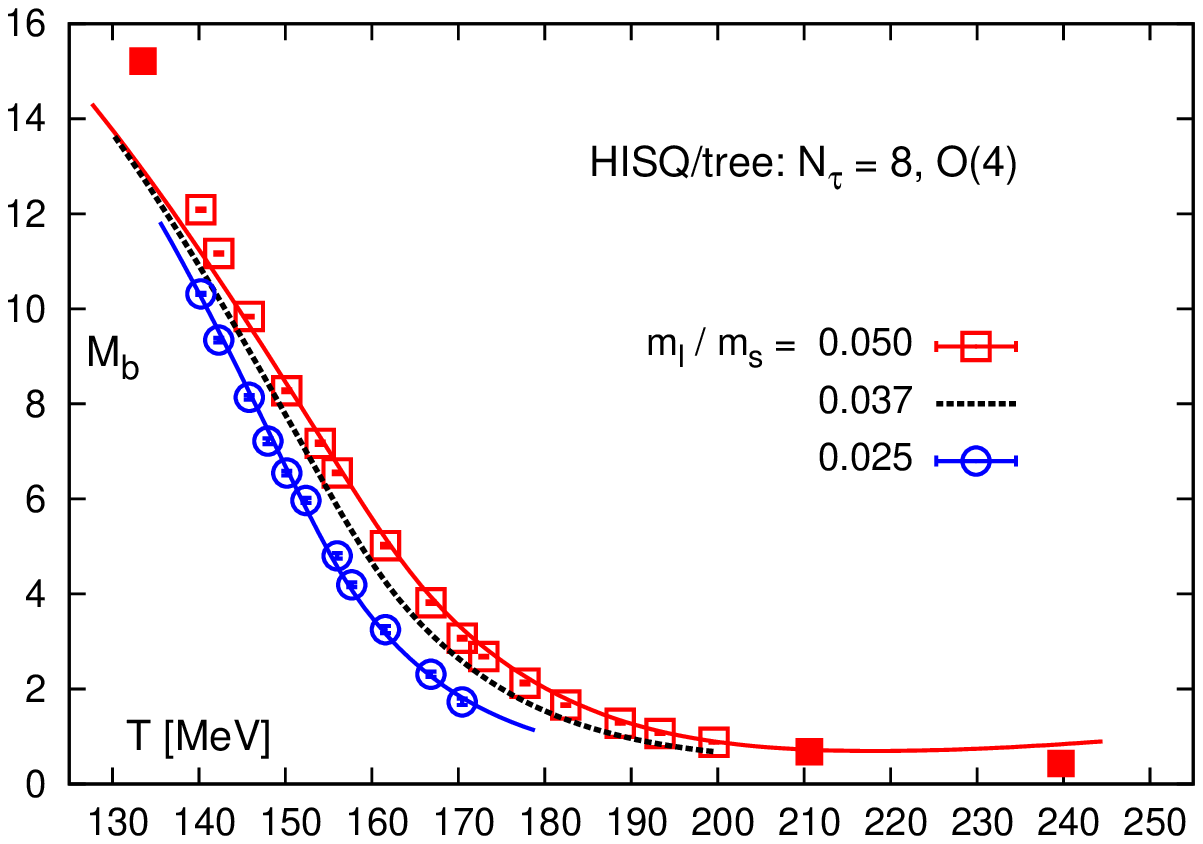}
\includegraphics[width=7cm]{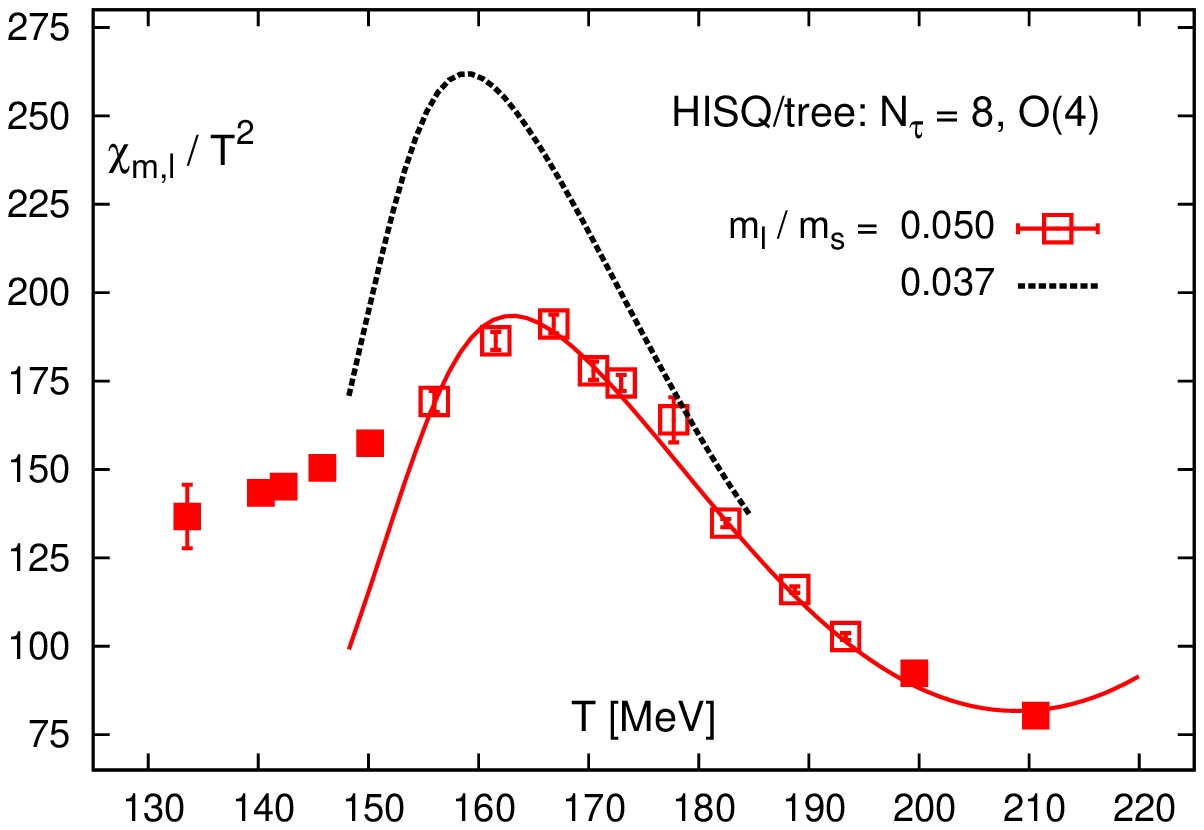}
\end{center}
\caption{
\label{fig:sfit_hisq8_O4} 
  Scaling fits and data for the chiral condensate $M_b$ calculated
  with the $HISQ/tree$ action on lattices with temporal extent
  $N_\tau=8$ (left) and the chiral susceptibility
  $\chi_{m,l}$(right). The data for $M_b$ at $m_l/m_s=0.025$ and for
  $M_b$ and $\chi_{m,l}$ at $m_l/m_s=0.05$ are fit simultaneously
  using the $O(4)$ scaling Ansatz. 
  The points used in the scaling fits are plotted using 
  open symbols.  The dotted lines give the data scaled to the physical
  quark masses.}
\end{figure}
\section{Temporal meson correlars and spectral functions}
Information on hadron properties at finite temperature as well as the transport coefficients 
are encoded in different spectral functions.
In particular the fate of different quarkonium states in QGP
can studied by calculating the corresponding quarkonium spectral functions 
(for a recent review see Ref. \cite{qgp09}).
On the lattice we can calculate correlation function in Euclidean time. 
This is related to the spectral function via integral
relation
\begin{equation}
G(\tau, T) = \int_0^{\infty} d \omega
\sigma(\omega,T) K(\tau,\omega,T) ,~~
K(\tau,\omega,T) = \frac{\cosh(\omega(\tau-1/2
T))}{\sinh(\omega/2 T)}.
\label{eq.kernel}
\end{equation}
Given the data on the Euclidean meson correlator $G(\tau, T)$ the meson spectral function 
can be calculated
using the Maximum Entropy Method (MEM)  \cite{mem}. For charmonium this was done by 
using correlators calculated on
isotropic lattices \cite{datta02,datta04} as well as  
anisotropic lattices \cite{umeda02,asakawa04,jako07} in the quenched approximation.
It has been found that quarkonium correlation function in Euclidean time show only very small temperature
dependence \cite{datta04,jako07}. In other channels, namely the vector, scalar and axial-vector channels 
stronger temperature dependence was found \cite{datta04,jako07}.
The spectral functions in the pseudo-scalar and vector channels reconstructed from MEM show peak 
structures which may
be interpreted as a ground state peak \cite{datta04,umeda02,asakawa04}. 
Together with the weak temperature dependence
of the correlation functions this was taken as strong indication that the 
1S charmonia ($\eta_c$ and $J/\psi$) survive
in the deconfined phase to temperatures as high as $1.6T_c$ \cite{datta04,umeda02,asakawa04}. A detailed study of
the systematic effects show, however, that the reconstruction of the charmonium spectral function is not reliable
at high temperatures \cite{jako07}, in particular the presence of peaks corresponding to bound states cannot be
reliably established. Presence of large cutoff effects 
at high frequencies also complicates the analysis \cite{freelat}.
The only statement that can be made is that the 
spectral function does not show significant changes 
within the errors of the calculations. 
Recently quarkonium spectral functions have been studied using potential models
and lattice data for the singlet free energy of static quark anti-quark pair \cite{mocsy3,mocsy4,mocsy5}. 
These calculations show that all
charmonium states are dissolved  at temperatures smaller than $1.2T_c$, but the Euclidean correlators do not show
significant changes and are in fairly good agreement with available lattice data  both 
for charmonium \cite{datta04,jako07} and bottomonium \cite{jako07,dattapanic05}. 
This is due to the fact that even in absence of bound states quarkonium spectral functions
show significant enhancement in the threshold region \cite{mocsy2}.  Therefore previous statements about quarkonia
survival at high temperatures have to be revisited. Exploratory calculations of the charmonium correlators and
spectral functions in 2-flavor QCD have been reported in Ref. \cite{aarts07} and the qualitative behavior
of the correlation functions was found to be similar.

The large enhancement of the quarkonium correlators above deconfinement in the scalar and axial-vector
channel can be understood in terms of the zero mode contribution \cite{mocsy2,umeda07} 
and not due to the dissolution of the $1P$ states as previously thought. 
Similar, though smaller in magnitude, enhancement of quarkonium correlators due to zero mode 
is seen also in the vector channel \cite{jako07}. Here it is related to heavy quark transport \cite{mocsy1,derek}.
Due to the heavy quark mass the Euclidean correlators for heavy quarkonium can be decomposed into
a high and low energy part $G(\tau,T)=G_{\rm low}(\tau,T)+G_{\rm high}(\tau,T)$
The area under the  peak in the spectral functions at zero energy $\omega \simeq 0$ 
giving the zero mode contribution
to the Euclidean correlator is proportional to some susceptibility, $G^i_{low}(\tau,T) \simeq T \chi^i(T)$, which
have been calculated on the lattice in Ref. \cite{me_hq08}. It is natural to ask whether 
the generalized susceptibilities can be described by a quasi-particle model. The generalized
susceptibilities have been calculated in Ref. \cite{aarts05} in the free theory.
Replacing the bare quark mass entering in  the expression of the generalized susceptibilities by an 
effective temperature dependent masses one can describe the zero mode contribution very 
well in all channels \cite{me_hq08}. 

While temporal correlators are not sensitive to the change in the spectral functions spatial quarkonium
correlation functions could be more sensitive to this. Recent lattice calculations show indication for
significant change in spatial charmonium correlators above deconfinement\cite{mukher}.

The analysis of the meson spectral functions was also performed in the light quark sector
\cite{ines01,karsch02,asakawaqm02,karschqm02}. Somewhat surprisingly this analysis revealed peak
structures in the deconfined phase which are difficult to interpret. Most likely the peak structures
are artifacts of the MEM analysis. The new analysis of the vector spectral function performed on fine
lattices and using several lattice spacings did not show any evidence for such structures \cite{okacz_dil}.
Furthermore, the same analysis revealed the expected transport peak at low frequencies and estimated
the electric conductivity to be \cite{okacz_dil}:
\begin{equation}
1/3 < \frac{\sigma}{C_{em} T} < 1,~~C_{em}=e^2 \sum_q Q_q^2.
\end{equation}

\section{Conclusions}

Recently significant progress has been made in lattice QCD calculations at non-zero temperature. Chiral
and deconfining aspects of the QCD transition have been studied using improved staggered quark formulation
allowing to control discretization effects. It has been shown that in the continuum limit different discretization
schemes give consistent results. In particular, agreement has been reached on the value of the chiral transition
temperature. There is still disagreement in the lattice calculation of the equation of state. Finally significant
progress has been made in lattice calculations of meson spectral functions that encode in-medium meson properties
and transport coefficients.

\section*{Acknowledgments} 
This work was supported by U.S. Department of Energy under
Contract No. DE-AC02-98CH10886.  Computations have been performed using
the USQCD resources on clusters at Fermilab and JLab and the QCDOC supercomputers
at BNL, as well as the BlueGene/L at the New York Center for Computational Sciences (NYCCS).


\begin{thebibliography}{00}  

\bibitem{gros81}
D. Gross, R. Pisarski, L. Yaffe, Rev. Mod. Phys. {\bf 53}, 43 (1981) 

\bibitem{nagle}
B. M\"uller and J. Nagle, Ann. Rev. Nucl. Part. Sci. {\bf 56}, 93 (2006) 

\bibitem{salgado}
U. Wiedemann, Nucl. Phys. A, {\bf 830}, 74c (2009)


\bibitem{kuti81}
%\cite{Kuti:1980gh}
%\bibitem{Kuti:1980gh}
  J.~Kuti, J.~Pol\'onyi and K.~Szlach\'anyi,
  %``Monte Carlo Study Of SU(2) Gauge Theory At Finite Temperature,''
  Phys.\ Lett.\  B {\bf 98} (1981) 199;
  %%CITATION = PHLTA,B98,199;%%

\bibitem{mclerran81}
%\cite{McLerran:1981pb}
%\bibitem{McLerran:1981pb}
  L.~D.~McLerran and B.~Svetitsky,
  %``Quark Liberation At High Temperature: A Monte Carlo Study Of SU(2) Gauge
  %Theory,''
  Phys.\ Rev.\  D {\bf 24} (1981) 450;
  %%CITATION = PHRVA,D24,450;%%

\bibitem{engels81}
%\cite{Engels:1980ty}
%\bibitem{Engels:1980ty}
  J.~Engels, F.~Karsch, H.~Satz and I.~Montvay,
  %``High Temperature SU(2) Gluon Matter On The Lattice,''
  Phys.\ Lett.\  B {\bf 101} (1981) 89.
  %%CITATION = PHLTA,B101,89;%%

\bibitem{boyd}
%\cite{Boyd:1996bx}
%\bibitem{Boyd:1996bx}
  G.~Boyd {\it et al.},
  %``Thermodynamics of SU(3) Lattice Gauge Theory,''
  Nucl.\ Phys.\  B {\bf 469}, 419 (1996)
%  [arXiv:hep-lat/9602007].
  %%CITATION = NUPHA,B469,419;%%

\bibitem{milc_old_eos}
C. W. Bernard et al., [MILC Collaboration], Phys. Rev. D {\bf 55} (1997) 6861
\bibitem{wilson_old}
A. Ali Khan et al., [CP-PACS Collaboration], 
Phys. Rev. D {\bf 64} (2001) 074510
\bibitem{p4_old}
%\cite{Karsch:2000ps}
%\bibitem{Karsch:2000ps}
  F.~Karsch, E.~Laermann and A.~Peikert,
  %``The pressure in 2, 2+1 and 3 flavour QCD,''
  Phys.\ Lett.\  B {\bf 478} (2000) 447; 
%  [arXiv:hep-lat/0002003].
  %%CITATION = PHLTA,B478,447;%%


\bibitem{milc_Tc}
  C.~Bernard {\it et al.}  [MILC Collaboration],
  %``QCD thermodynamics with three flavors of improved staggered quarks,''
  Phys.\ Rev.\ D {\bf 71}, 034504 (2005)

\bibitem{fodor05}
%\cite{Aoki:2005vt}
%\bibitem{Aoki:2005vt}
  Y.~Aoki {\it et al.}
  %``The equation of state in lattice QCD: With physical quark masses  towards
  %the continuum limit,''
  JHEP {\bf 0601}, 089 (2006)
%  [arXiv:hep-lat/0510084].
  %%CITATION = HEP-LAT 0510084;%%

\bibitem{our_Tc}
%\cite{Cheng:2006qk}
%\bibitem{Cheng:2006qk}
  M.~Cheng {\it et al.},
  %``The transition temperature in QCD,''
  Phys.\ Rev.\  D {\bf 74}, 054507 (2006)
 % [arXiv:hep-lat/0608013].
  %%CITATION = PHRVA,D74,054507;%%

\bibitem{fodor06}
%%\cite{Aoki:2006br}
%\bibitem{Aoki:2006br}
  Y.~Aoki {\it et al.},
  %``The QCD transition temperature: Results with physical masses in the
  %continuum limit,''
  Phys.\ Lett.\  B {\bf 643} (2006) 46
  [arXiv:hep-lat/0609068].
  %%CITATION = PHLTA,B643,46;%%

\bibitem{milc_eos}
%\cite{Bernard:2006nj}
%\bibitem{Bernard:2006nj}
  C.~Bernard {\it et al.},
  %``QCD equation of state with 2+1 flavors of improved staggered quarks,''
  Phys.\ Rev.\  D {\bf 75} (2007) 094505
 % [arXiv:hep-lat/0611031].
  %%CITATION = PHRVA,D75,094505;%%

\bibitem{our_eos}
%\cite{Cheng:2007jq}
%\bibitem{Cheng:2007jq}
  M.~Cheng {\it et al.},
  %``The QCD Equation of State with almost Physical Quark Masses,''
  Phys.\ Rev.\  D {\bf 77} (2008) 014511
%  [arXiv:0710.0354 [hep-lat]].
  %%CITATION = PHRVA,D77,014511;%%

\bibitem{eos005}
%\cite{Cheng:2009zi}
%\bibitem{Cheng:2009zi}
  M.~Cheng {\it et al.},
  %``Equation of State for physical quark masses,''
  Phys.\ Rev.\  D {\bf 81}, 054504 (2010)
%  [arXiv:0911.2215 [hep-lat]].
  %%CITATION = PHRVA,D81,054504;%%

\bibitem{hotQCD}
%\cite{Bazavov:2009zn}
%\bibitem{Bazavov:2009zn}
  A.~Bazavov {\it et al.},
  %``Equation of state and QCD transition at finite temperature,''
  Phys.\ Rev.\  D {\bf 80} (2009), 014504
%  [arXiv:0903.4379 [hep-lat]].
  %%CITATION = ARXIV:0903.4379;%%

\bibitem{fodor09}
%\cite{Aoki:2009sc}
%\bibitem{Aoki:2009sc}
  Y.~Aoki, {\it et al},
  %``The QCD transition temperature: results with physical masses in the
  %continuum limit II,''
  JHEP {\bf 0906} (2009) 088
%  [arXiv:0903.4155 [hep-lat]].
  %%CITATION = JHEPA,0906,088;%%

\bibitem{fodor10}
%\cite{Borsanyi:2010bp}
%\bibitem{Borsanyi:2010bp} 
  S.~Borsanyi {\it et al.}  [Wuppertal-Budapest Collaboration],
  %``Is there still any T_c mystery in lattice QCD? Results with physical masses in the continuum limit III,''
  JHEP {\bf 1009}, 073 (2010)
%  [arXiv:1005.3508 [hep-lat]].
  %%CITATION = ARXIV:1005.3508;%%


\bibitem{fodor10eos}
%\cite{Borsanyi:2010cj}
%\bibitem{Borsanyi:2010cj} 
  S.~Borsanyi {\it et al},
  %``The QCD equation of state with dynamical quarks,''
  JHEP {\bf 1011}, 077 (2010)
%  [arXiv:1007.2580 [hep-lat]].
  %%CITATION = ARXIV:1007.2580;%%


\bibitem{wwnd10}
%\cite{Bazavov:2010sb}
%\bibitem{Bazavov:2010sb}
  A.~Bazavov and P.~Petreczky  [HotQCD collaboration],
  %``Deconfinement and chiral transition with the highly improved staggered
  %quark (HISQ) action,''
  J.\ Phys.\ Conf.\ Ser.\  {\bf 230}, 012014 (2010);
%  [arXiv:1005.1131 [hep-lat]].
  %%CITATION = 00462,230,012014;%%

\bibitem{lat09}
%\cite{Bazavov:2009mi}
%\bibitem{Bazavov:2009mi}
A.~Bazavov and P.~Petreczky  [HotQCD Collaboration],
  %``First results on QCD thermodynamics with HISQ action,''
  PoS {\bf LAT2009}, 163 (2009)
%  [arXiv:0912.5421 [hep-lat]].
  %%CITATION = POSCI,LAT2009,163;%%

\bibitem{hotqcd2}
%\cite{Bazavov:2011nk}
%\bibitem{Bazavov:2011nk} 
  A.~Bazavov {\it et al.},
  %``The chiral and deconfinement aspects of the QCD transition,''
  arXiv:1111.1710 [hep-lat].
  %%CITATION = ARXIV:1111.1710;%%


\bibitem{prasad}
P. Hegde, arXiv:1112.0364v1 [hep-lat]

\bibitem{fodor11}
%\cite{Borsanyi:2011kg}
%\bibitem{Borsanyi:2011kg} 
  S.~Borsanyi {\it et al.},
  %``QCD thermodynamics with Wilson fermions,''
  arXiv:1111.3500 [hep-lat].
  %%CITATION = ARXIV:1111.3500;%%


\bibitem{whotqcd}
%\cite{Umeda:2010ye}
%%%%%\bibitem{Umeda:2010ye} 
  T.~Umeda {\it et al.}  [WHOT-QCD Collaboration],
  %``EOS in 2+1 flavor QCD with improved Wilson quarks by the fixed-scale approach,''
  PoS LATTICE {\bf 2010}, 218 (2010);
% [arXiv:1011.2548 [hep-lat]].
  %%CITATION = ARXIV:1011.2548;%%
%\cite{Ejiri:2011kp}
%\bibitem{Ejiri:2011kp} 
%  S.~Ejiri {\it et al.}  [WHOT-QCD Collaboration],
  %``Scaling behavior of chiral phase transition in two-flavor QCD with improved Wilson quarks at finite density,''
  arXiv:1101.5582 [hep-lat].
  %%CITATION = ARXIV:1101.5582;%%



\bibitem{orginos}
%\cite{Orginos:1999cr}
%\bibitem{Orginos:1999cr}
  K.~Orginos, D.~Toussaint and R.~L.~Sugar  [MILC Collaboration],
  %``Variants of fattening and flavor symmetry restoration,''
  Phys.\ Rev.\  D {\bf 60} (1999) 054503
 % [arXiv:hep-lat/9903032].
  %%CITATION = PHRVA,D60,054503;%%

\bibitem{latproc}
%\cite{Bazavov:2010pg}
%\bibitem{Bazavov:2010pg} 
  A.~Bazavov {\it et al.}  [for the HotQCD Collaboration],
  %``Taste symmetry and QCD thermodynamics with improved staggered fermions,''
  PoS LATTICE {\bf 2010}, 169 (2010);
%  [arXiv:1012.1257 [hep-lat]].
  %%CITATION = ARXIV:1012.1257;%%
%\cite{Soldner:2010xk}
%\bibitem{Soldner:2010xk} 
  W.~Soldner [HotQCD Collaboration],
  %``Chiral Aspects of Improved Staggered Fermions with 2+1-Flavors from the HotQCD Collaboration,''
  PoS LATTICE {\bf 2010}, 215 (2010)
%  [arXiv:1012.4484 [hep-lat]].
  %%CITATION = ARXIV:1012.4484;%%

\bibitem{fodor04}
%\cite{Csikor:2004ik}
%\bibitem{Csikor:2004ik}
  F.~Csikor, {\it et al.},
  %``Equation of state at finite temperature and chemical potential, lattice
  %QCD results,''
  JHEP {\bf 0405} (2004) 046
%  [arXiv:hep-lat/0401016].
  %%CITATION = JHEPA,0405,046;%%

\bibitem{pasi}
%\cite{Huovinen:2009yb}
%\bibitem{Huovinen:2009yb}
  P.~Huovinen and P.~Petreczky,
  %``QCD Equation of State and Hadron Resonance Gas,''
  Nucl.\ Phys.\  A {\bf 837}, 26 (2010)
%  [arXiv:0912.2541 [hep-ph]].
  %%CITATION = NUPHA,A837,26;%%


\bibitem{blaizot}
%\cite{Blaizot:1999ip}
%\bibitem{Blaizot:1999ip}
  J.~P.~Blaizot, E.~Iancu and A.~Rebhan,
  %``The entropy of the QCD plasma,''
  Phys.\ Rev.\ Lett.\  {\bf 83} (1999) 2906; 
%  [arXiv:hep-ph/9906340].
  %%CITATION = PRLTA,83,2906;%%

\bibitem{blaizot1}
%\cite{Blaizot:2000fc}
%\bibitem{Blaizot:2000fc}
J.~P.~Blaizot, E.~Iancu and A.~Rebhan,
  %``Approximately self-consistent resummations for the thermodynamics of  the
  %quark-gluon plasma. I: Entropy and density,''
  Phys.\ Rev.\  D {\bf 63} (2001) 065003
%  [arXiv:hep-ph/0005003].
  %%CITATION = PHRVA,D63,065003;%%
\bibitem{gubser98}
S.S. Gubser, I.R. Klebanov and I.I Tseytlin, Nucl. Phys. B {\bf 534} (1998) 202

\bibitem{mike}
%\cite{Andersen:2010wu}
%\bibitem{Andersen:2010wu}
  J.~O.~Andersen, L.~E.~Leganger, M.~Strickland and N.~Su,
  %``NNLO hard-thermal-loop thermodynamics for QCD,''
  arXiv:1009.4644 [hep-ph].
  %%CITATION = ARXIV:1009.4644;%%

\bibitem{our_fluct}
%\cite{Cheng:2008zh}
%\bibitem{Cheng:2008zh}
  M.~Cheng {\it et al.},
  %``Baryon Number, Strangeness and Electric Charge Fluctuations in QCD at High
  %Temperature,''
  Phys.\ Rev.\  D {\bf 79} (2009) 074505
%  [arXiv:0811.1006 [hep-lat]].
  %%CITATION = PHRVA,D79,074505;%%

\bibitem{fodor_fluct}
%\cite{Borsanyi:2011bm}
%\bibitem{Borsanyi:2011bm} 
  S.~Borsanyi {\it et al.}  [Wuppertal-Budapest Collaboration],
  %``Correlations and fluctuations from lattice QCD,''
  J.\ Phys.\ G G {\bf 38}, 124060 (2011)
%  [arXiv:1109.5030 [hep-lat]].
  %%CITATION = ARXIV:1109.5030;%%


\bibitem{blaizot01}
%\cite{Blaizot:2001vr}
%\bibitem{Blaizot:2001vr}
  J.~P.~Blaizot, E.~Iancu and A.~Rebhan,
  %``Quark number susceptibilities from HTL-resummed thermodynamics,''
  Phys.\ Lett.\  B {\bf 523} (2001) 143
%  [arXiv:hep-ph/0110369].
  %%CITATION = PHLTA,B523,143;%%

\bibitem{rebhan_sewm02}
%\cite{Rebhan:2003fj}
%\bibitem{Rebhan:2003fj}
  A.~Rebhan,
  %``HTL-resummed thermodynamics of hot and dense QCD: An update,''
  arXiv:hep-ph/0301130.
  %%CITATION = HEP-PH/0301130;%%

\bibitem{munshi}
%\cite{Haque:2010rb}
%\bibitem{Haque:2010rb}
  N.~Haque and M.~G.~Mustafa,
  %``Quark Number Susceptibility in Hard Thermal Loop Perturbation Theory,''
  arXiv:1007.2076 [hep-ph].
  %%CITATION = ARXIV:1007.2076;%%

%\bibitem{my_qm09}
%\cite{Petreczky:2009at}
%\bibitem{Petreczky:2009at} 
%  P.~Petreczky,
  %``Lattice QCD at finite temperature : Present status,''
%  Nucl.\ Phys.\ A {\bf 830}, 11C (2009)
%  [arXiv:0908.1917 [hep-ph]].
  %%CITATION = ARXIV:0908.1917;%%


\bibitem{progress}
%\cite{Petreczky:2009cr}
%\bibitem{Petreczky:2009cr}
  P.~Petreczky, P.~Hegde and A.~Velytsky  [RBC-Bielefeld Collaboration],
  %``Quark number fluctuations at high temperatures,''
  PoS {\bf LAT2009}, 159 (2009)
%  [arXiv:0911.0196 [hep-lat]].
  %%CITATION = POSCI,LAT2009,159;%%

\bibitem{vs1}
%\cite{Kim:2009uu}
%\bibitem{Kim:2009uu} 
  K.~-Y.~Kim and J.~Liao,
  %``On the Baryonic Density and Susceptibilities in a Holographic Model of QCD,''
  Nucl.\ Phys.\ B {\bf 822}, 201 (2009)
%  [arXiv:0906.2978 [hep-th]].
  %%CITATION = ARXIV:0906.2978;%%

\bibitem{vs2}
%\cite{Kim:2010zg}
%\bibitem{Kim:2010zg} 
  Y.~Kim, Y.~Matsuo, W.~Sim, S.~Takeuchi and T.~Tsukioka,
  %``Quark Number Susceptibility with Finite Chemical Potential in Holographic QCD,''
  JHEP {\bf 1005}, 038 (2010)
%  [arXiv:1001.5343 [hep-th]].
  %%CITATION = ARXIV:1001.5343;%%

\bibitem{mocsy1}
%\cite{Petreczky:2010tk}
%\bibitem{Petreczky:2010tk} 
  P.~Petreczky, C.~Miao and A.~Mocsy,
  %``Quarkonium spectral functions with complex potential,''
  Nucl.\ Phys.\ A {\bf 855}, 125 (2011)
% [arXiv:1012.4433 [hep-ph]].
  %%CITATION = ARXIV:1012.4433;%%


\bibitem{mocsy2}
%\cite{Mocsy:2005qw}
%\bibitem{Mocsy:2005qw}
\'A.~M\'ocsy and P.~Petreczky,
  %``Quarkonia correlators above deconfinement,''
  Phys.\ Rev.\  D {\bf 73}, 074007 (2006)
%  [arXiv:hep-ph/0512156].
  %%CITATION = PHRVA,D73,074007;%%

\bibitem{mocsy3}
%\cite{Mocsy:2007jz}
%\bibitem{Mocsy:2007jz}
\'A.~M\'ocsy and P.~Petreczky,
  %``Color Screening Melts Quarkonium,''
  Phys.\ Rev.\ Lett.\  {\bf 99}, 211602 (2007);
%  [arXiv:0706.2183 [hep-ph]].
  %%CITATION = PRLTA,99,211602;%%


\bibitem{mocsy4}
%\cite{Mocsy:2007yj}
%\bibitem{Mocsy:2007yj}
\'A.~M\'ocsy and P.~Petreczky,
  %``Can quarkonia survive deconfinement ?,''
  Phys.\ Rev.\  D {\bf 77}, 014501 (2008);
%  [arXiv:0705.2559 [hep-ph]].
  %%CITATION = PHRVA,D77,014501;%%

\bibitem{mocsy5}
%\bibitem{Mocsy:2007py}
\'A.~M\'ocsy and P.~Petreczky,
  %``Describing Charmonium Correlation Functions in Euclidean Time,''
  Eur.\ Phys.\ J.\ ST {\bf 155}, 101 (2008)
%  [arXiv:0710.5125 [hep-ph]].
  %%CITATION = 00619,155,101;%%

\bibitem{rbc_f1}
%\cite{Petreczky:2010yn}
%\bibitem{Petreczky:2010yn} 
  P.~Petreczky,
  %``Quarkonium in Hot Medium,''
  J.\ Phys.\ G  {\bf 37}, 094009 (2010)
%  [arXiv:1001.5284 [hep-ph]].
  %%CITATION = ARXIV:1001.5284;%%


\bibitem{kostya1}
%\cite{Petreczky:2004pz}
%\bibitem{Petreczky:2004pz}
  P.~Petreczky and K.~Petrov,
  %``Free energy of a static quark anti-quark pair and the renormalized
  %Polyakov loop in three flavor QCD,''
  Phys.\ Rev.\  D {\bf 70}, 054503 (2004)
%  [arXiv:hep-lat/0405009].
  %%CITATION = PHRVA,D70,054503;%%

\bibitem{okacz02}
O.~Kaczmarek {\it et al.},
%  O.~Kaczmarek, F.~Karsch, P.~Petreczky and F.~Zantow,
  %``Heavy Quark Anti-Quark Free Energy and the Renormalized Polyakov Loop,''
  Phys.\ Lett.\  B {\bf 543}, 41 (2002)
%  [arXiv:hep-lat/0207002].
  %%CITATION = PHLTA,B543,41;%%


\bibitem{speculations}
%\cite{Hatta:2003bc}
%\bibitem{Hatta:2003bc}
 see e.g.  Y.~Hatta and K.~Fukushima,
  %``On the nature of thermal QCD phase transitions,''
  arXiv:hep-ph/0311267, and references therein.
  %%CITATION = HEP-PH/0311267;%%

\bibitem{digal03}
%\cite{Digal:2003jc}
%\bibitem{Digal:2003jc}
  S.~Digal, S.~Fortunato and P.~Petreczky,
  %``Heavy quark free energies and screening in SU(2) gauge theory,''
  Phys.\ Rev.\  D {\bf 68}, 034008 (2003)
%  [arXiv:hep-lat/0304017].
  %%CITATION = PHRVA,D68,034008;%%


\bibitem{okacz05}
%\cite{Kaczmarek:2005ui}
%\bibitem{Kaczmarek:2005ui}
  O.~Kaczmarek and F.~Zantow,
  %``Static quark anti-quark interactions in zero and finite temperature  QCD.
  %I: Heavy quark free energies, running coupling and quarkonium  binding,''
  Phys.\ Rev.\  D {\bf 71}, 114510 (2005)
 % [arXiv:hep-lat/0503017].
  %%CITATION = PHRVA,D71,114510;%


\bibitem{mehard04}
%\cite{Petreczky:2005bd}
%\bibitem{Petreczky:2005bd}
  P.~Petreczky,
  %``Heavy quark potentials and quarkonia binding,''
  Eur.\ Phys.\ J.\  C {\bf 43}, 51 (2005)
 % [arXiv:hep-lat/0502008].
  %%CITATION = EPHJA,C43,51;%%

\bibitem{qgp09}
%\cite{Bazavov:2009us}
%\bibitem{Bazavov:2009us}
  A.~Bazavov, P.~Petreczky and A.~Velytsky,
  %``Quarkonium at Finite Temperature,''
  arXiv:0904.1748 [hep-ph].
  %%CITATION = ARXIV:0904.1748;%%


\bibitem{baza08}
%\cite{Bazavov:2008rw}
%\bibitem{Bazavov:2008rw}
  A.~Bazavov, P.~Petreczky and A.~Velytsky,
  %``Static quark anti-quark pair in SU(2) gauge theory,''   
  Phys.\ Rev.\  D {\bf 78}, 114026 (2008)
%  [arXiv:0809.2062 [hep-lat]].
  %%CITATION = PHRVA,D78,114026;%%

\bibitem{plc_new}
%\cite{Brambilla:2010xn}
%\bibitem{Brambilla:2010xn} 
  N.~Brambilla {\it et al.},
  %``The Polyakov loop and correlator of Polyakov loops at next-to-next-to-leading order,''
  Phys.\ Rev.\ D {\bf 82}, 074019 (2010)
%  [arXiv:1007.5172 [hep-ph]].
  %%CITATION = ARXIV:1007.5172;%%


\bibitem{sigma_s}
%\cite{Cheng:2008bs}
%\bibitem{Cheng:2008bs} 
  M.~Cheng {\it et al.},
  %``The Spatial String Tension and Dimensional Reduction in QCD,''
  Phys.\ Rev.\ D {\bf 78}, 034506 (2008)
%  [arXiv:0806.3264 [hep-lat]].
  %%CITATION = ARXIV:0806.3264;%%


\bibitem{prop98}
%\cite{Karsch:1998tx}
%\bibitem{Karsch:1998tx} 
  F.~Karsch, M.~Oevers and P.~Petreczky,
  %``Screening masses of hot SU(2) gauge theory from the 3-d adjoint Higgs model,''
  Phys.\ Lett.\ B {\bf 442}, 291 (1998)
%  [hep-lat/9807035].
  %%CITATION = HEP-LAT/9807035;%%



\bibitem{prop00}
%\cite{Cucchieri:2000cy}
%\bibitem{Cucchieri:2000cy} 
  A.~Cucchieri, F.~Karsch and P.~Petreczky,
  %``Magnetic screening in hot nonAbelian gauge theory,''
  Phys.\ Lett.\ B {\bf 497}, 80 (2001)
%  [hep-lat/0004027].
  %%CITATION = HEP-LAT/0004027;%%


\bibitem{prop01}
%\cite{Cucchieri:2001tw}
%\bibitem{Cucchieri:2001tw} 
  A.~Cucchieri, F.~Karsch and P.~Petreczky,
  %``Propagators and dimensional reduction of hot SU(2) gauge theory,''
  Phys.\ Rev.\ D {\bf 64}, 036001 (2001)
%  [hep-lat/0103009].
  %%CITATION = HEP-LAT/0103009;%%



\bibitem{MEoS}
%\cite{Ejiri:2009ac}
%\bibitem{Ejiri:2009ac} 
  S.~Ejiri {\it et al.},
  %``On the magnetic equation of state in (2+1)-flavor QCD,''
  Phys.\ Rev.\ D {\bf 80}, 094505 (2009)
%  [arXiv:0909.5122 [hep-lat]].
  %%CITATION = ARXIV:0909.5122;%%


\bibitem{nature}
%\cite{Aoki:2006we}
%\bibitem{Aoki:2006we}
  Y.~Aoki, {\it et al.},
  %``The order of the quantum chromodynamics transition predicted by the
  %standard model of particle physics,''
  Nature {\bf 443}, 675 (2006)
 % [arXiv:hep-lat/0611014].
  %%CITATION = NATUA,443,675;%%

\bibitem{pisarski81}
R.D. Pisarski and F. Wilczek, Phys. Rev. D {\bf 49}, 338 (1984)

\bibitem{mscr}
%\cite{Cheng:2010fe}
%\bibitem{Cheng:2010fe} 
  M.~Cheng {\it et al.},
  %``Meson screening masses from lattice QCD with two light and the strange quark,''
  Eur.\ Phys.\ J.\ C {\bf 71}, 1564 (2011)
%  [arXiv:1010.1216 [hep-lat]].
  %%CITATION = ARXIV:1010.1216;%%


\bibitem{curvature}
%\cite{Kaczmarek:2011zz}
%\bibitem{Kaczmarek:2011zz} 
  O.~Kaczmarek {\it et al.},
  %``Phase boundary for the chiral transition in (2+1) -flavor QCD at small values of the chemical potential,''
  Phys.\ Rev.\ D {\bf 83}, 014504 (2011)
%  [arXiv:1011.3130 [hep-lat]].
  %%CITATION = ARXIV:1011.3130;%%


\bibitem{EngelsO2}
  J.~Engels, S.~Holtmann, T.~Mendes and T.~Schulze,
  %``Equation of state and Goldstone-mode effects of the three-dimensional  O(2)
  %model,''
  Phys.\ Lett.\  B {\bf 492}, 219 (2000).
\bibitem{Toussaint}
  D.~Toussaint,
  %``Scaling functions for O(4) in three dimensions,''
  Phys.\ Rev.\  D {\bf 55}, 362 (1997).
\bibitem{EngelsO4}
J. Engels and T. Mendes,
  %``Goldstone-mode effects and scaling function for the three-dimensional  O(4)
  %model,''
  Nucl.\ Phys.\  B {\bf 572}, 289 (2000).
\bibitem{Engels2001}
  J.~Engels, S.~Holtmann, T.~Mendes and T.~Schulze,
  %``Finite-size-scaling functions for 3d O(4) and O(2) spin models and QCD,''
  Phys.\ Lett.\  B {\bf 514}, 299 (2001).

\bibitem{mem}
%\cite{Asakawa:2000tr}
%\bibitem{Asakawa:2000tr}
  M.~Asakawa, T.~Hatsuda and Y.~Nakahara,
  %``Maximum entropy analysis of the spectral functions in lattice QCD,''
  Prog.\ Part.\ Nucl.\ Phys.\  {\bf 46}, 459 (2001)
%  [arXiv:hep-lat/0011040].
  %%CITATION = HEP-LAT 0011040;%%

\bibitem{datta02}
%\cite{Datta:2002ck}
%\bibitem{Datta:2002ck}
  S.~Datta et al.,
  %``A study of charmonium systems across the deconfinement transition,''
  Nucl.\ Phys.\ Proc.\ Suppl.\  {\bf 119}, 487 (2003)
%  [arXiv:hep-lat/0208012].
  %%CITATION = HEP-LAT 0208012;%%

\bibitem {datta04}
S.~Datta, {\it et al.,}
%``Behavior of charmonium systems after deconfinement,''
Phys.\ Rev.\ D \textbf{69}, 094507 (2004)
%[arXiv:hep-lat/0312037].
%%CITATION = HEP-LAT 0312037;%%

\bibitem {umeda02}
%\cite{Umeda:2002vr}
%\bibitem{Umeda:2002vr}
  T.~Umeda, K.~Nomura and H.~Matsufuru,
  %``Charmonium at finite temperature in quenched lattice QCD,''
  Eur.\ Phys.\ J.\  C {\bf 39S1}, 9 (2005)
 % [arXiv:hep-lat/0211003].
  %%CITATION = EPHJA,C39S1,9;%%


\bibitem {asakawa04}
M.~Asakawa and T.~Hatsuda,
%``J/psi and eta/c in the deconfined plasma from lattice QCD,''
Phys.\ Rev.\ Lett.\ \textbf{92}, 012001 (2004)
%%CITATION = HEP-LAT 0308034;%%

\bibitem{jako07}
%\cite{Jakovac:2006sf}
%\bibitem{Jakovac:2006sf}
  A.~Jakov\'ac {\it et al.},
  %``Quarkonium correlators and spectral functions at zero and finite
  %temperature,''
  Phys.\ Rev.\  D {\bf 75}, 014506 (2007)
 % [arXiv:hep-lat/0611017].
  %%CITATION = PHRVA,D75,014506;%%

\bibitem{dattapanic05}
%\cite{Datta:2006ua}
%\bibitem{Datta:2006ua}
  S.~Datta {\it et al.,}
  %``Quarkonia in a deconfined gluonic plasma,''
  AIP Conf.\ Proc.\  {\bf 842}, 35 (2006)
 % [arXiv:hep-lat/0603002].
  %%CITATION = APCPC,842,35;%%

\bibitem{aarts07}
%\cite{Aarts:2007pk}
%\bibitem{Aarts:2007pk}
  G.~Aarts {\it et al.,}
  %``Charmonium at high temperature in two-flavor QCD,''
  Phys.\ Rev.\  D {\bf 76}, 094513 (2007)
 % [arXiv:0705.2198 [hep-lat]].
  %%CITATION = PHRVA,D76,094513;%%


\bibitem{freelat}
%\cite{Karsch:2003wy}
%\bibitem{Karsch:2003wy}
  F.~Karsch, {\it et al.,}
  %``Infinite temperature limit of meson spectral functions calculated on  the
  %lattice,''
  Phys.\ Rev.\  D {\bf 68}, 014504 (2003)
  [arXiv:hep-lat/0303017].
  %%CITATION = PHRVA,D68,014504;%%


\bibitem{umeda07}
%\cite{Umeda:2007hy}
%\bibitem{Umeda:2007hy}
  T.~Umeda,
  %``A constant contribution in meson correlators at finite temperature,''
  Phys.\ Rev.\  D {\bf 75}, 094502 (2007)
%  [arXiv:hep-lat/0701005].
  %%CITATION = PHRVA,D75,094502;%%

\bibitem{derek}
%\cite{Petreczky:2005nh}
%\bibitem{Petreczky:2005nh}
P.~Petreczky and D.~Teaney,
%``Heavy quark diffusion from the lattice,''
Phys.\ Rev.\  D {\bf 73}, 014508 (2006)
% [arXiv:hep-ph/0507318].
%%CITATION = PHRVA,D73,014508;%%

\bibitem{me_hq08}
%\cite{Petreczky:2008px}
%\bibitem{Petreczky:2008px}
  P.~Petreczky,
  %``On temperature dependence of quarkonium correlators,''
  Eur.\ Phys.\ J.\  C {\bf 62}, 85 (2009)
%  [arXiv:0810.0258 [hep-lat]].
  %%CITATION = EPHJA,C62,85;%%

\bibitem{aarts05}
%\cite{Aarts:2005hg}
%\bibitem{Aarts:2005hg}
  G.~Aarts and J.~M.~Martinez Resco,
  %``Continuum and lattice meson spectral functions at nonzero momentum and
  %high temperature,''
  Nucl.\ Phys.\  B {\bf 726} (2005) 93
%  [arXiv:hep-lat/0507004].
  %%CITATION = NUPHA,B726,93;%%

\bibitem{mukher}
%\cite{Mukherjee:2008tr}
%\bibitem{Mukherjee:2008tr}
  S.~Mukherjee,
  %``Screening of light mesons and charmonia at high temperature,''
  Nucl.\ Phys.\  A {\bf 820} (2009) 283C
%  [arXiv:0810.2906 [hep-lat]].
  %%CITATION = NUPHA,A820,283C;%%


\bibitem{ines01}
%\cite{Wetzorke:2001dk}
%\bibitem{Wetzorke:2001dk} 
  I.~Wetzorke, F.~Karsch, E.~Laermann, P.~Petreczky and S.~Stickan,
  %``Meson spectral functions at finite temperature,''
  Nucl.\ Phys.\ Proc.\ Suppl.\  {\bf 106}, 510 (2002)
%  [hep-lat/0110132].
 %%CITATION = HEP-LAT/0110132;%%


\bibitem{karsch02}
%\cite{Karsch:2001uw}
%\bibitem{Karsch:2001uw}
  F.~Karsch {\it et al.,}
  %``A lattice calculation of thermal dilepton rates,''
  Phys.\ Lett.\  B {\bf 530}, 147 (2002);
 % [arXiv:hep-lat/0110208].
  %%CITATION = PHLTA,B530,147;%%

\bibitem{karschqm02}
 F.~Karsch {\it et al.,}
%\bibitem{Karsch:2002wv}
%  F.~Karsch, S.~Datta, E.~Laermann, P.~Petreczky, S.~Stickan and I.~Wetzorke,
  %``Hadron correlators, spectral functions and thermal dilepton rates from
  %lattice QCD,''
  Nucl.\ Phys.\  A {\bf 715}, 701 (2003)
%  [arXiv:hep-ph/0209028].
  %%CITATION = NUPHA,A715,701;%%


\bibitem{asakawaqm02}
%\cite{Asakawa:2002xj}
%\bibitem{Asakawa:2002xj}
  M.~Asakawa, T.~Hatsuda and Y.~Nakahara,
  %``Hadronic spectral functions above the QCD phase transition,''
  Nucl.\ Phys.\  A {\bf 715}, 863 (2003)
 % [Nucl.\ Phys.\ Proc.\ Suppl.\  {\bf 119}, 481 (2003)]
 % [arXiv:hep-lat/0208059].
  %%CITATION = NUPHZ,119,481;%%

\bibitem{okacz_dil}
%\bibitem{Ding:2010ga} 
  H.~-T.~Ding {\it et al.,}
  %``Thermal dilepton rate and electrical conductivity: An analysis of vector current correlation functions in quenched lattice QCD,''
  Phys.\ Rev.\ D {\bf 83}, 034504 (2011)
%  [arXiv:1012.4963 [hep-lat]].
  %%CITATION = ARXIV:1012.4963;%%

\end{thebibliography}
\end{document}